\documentclass[camera,letterpaper,nomarginnotes,nonarrowgutter]{jpaper}
\usepackage{amsmath,amssymb,amsfonts}
\usepackage{tikz}
\usepackage{mathptmx} % This is Times font
\usepackage{comment}
\usepackage{algorithm}
\usepackage{graphics}
\usepackage{fancyhdr}
\usepackage{booktabs}
\usepackage{pifont}
\newcommand{\cmark}{\ding{51}}%
\newcommand{\xmark}{\ding{55}}%
\usepackage[normalem]{ulem}
\usepackage[sort,compress]{cite}
\usepackage{hhline}
\usepackage{tikz}
\usepackage{xcolor}
\usepackage{setspace}
\usepackage{textcomp}
\usepackage{hhline}
\usepackage{float}
\usepackage{afterpage}
\usepackage{graphicx}
\usepackage{hyphenat}
\usepackage{makecell}
\usepackage{xspace}
\usepackage{listings}
\usepackage{multirow}
\usepackage{balance}
\usepackage{caption}
\usepackage{algpseudocode}
\usepackage{multicol}
\usepackage{hyperref}
\usepackage{duckuments} %duck figures
\usepackage{datetime}
\usepackage{datetime2}

\def\BibTeX{{\rm B\kern-.05em{\sc i\kern-.025em b}\kern-.08em
    T\kern-.1667em\lower.7ex\hbox{E}\kern-.125emX}}

\usepackage{xspace}
\usepackage[acronym,nonumberlist,nowarn]{glossaries}
    \glsdisablehyper
     \newacronym{pim}{PiM}{Processing-in-Memory}
\newacronym{pid}{PiD}{Processing-in-DRAM}
\newacronym{pnm}{PnM}{Processing-near-Memory}
\newacronym{pnd}{PnD}{Processing-near-DRAM}
\newacronym{pum}{PuM}{Processing-using-Memory}
\newacronym{pud}{PuD}{Processing-using-DRAM}

\newcommand{\X}[0]{IMPACT}

\definecolor{darkspringgreen}{rgb}{0.09, 0.45, 0.27}
\definecolor{denim}{rgb}{0.08, 0.38, 0.74}
\definecolor{darkolivegreen}{rgb}{0.33, 0.42, 0.18}
\definecolor{tangerine}{rgb}{0.95, 0.52, 0.0}
\definecolor{mahogany}{rgb}{0.75, 0.25, 0.0}
\definecolor{nbs}{rgb}{0.88, 0.07, 0.37}
\definecolor{cornflowerblue}{rgb}{0.39, 0.58, 0.93}
\definecolor{darkpink}{rgb}{0.88, 0.28, 0.54}
\definecolor{forestgreen}{rgb}{0.0, 0.27, 0.13}
\definecolor{amber}{rgb}{1.0, 0.49, 0.0}
\definecolor{azure(colorwheel)}{rgb}{0.0, 0.5, 1.0}
\definecolor{greenp}{rgb}{0.0, 0.65, 0.50}
\definecolor{peach}{rgb}{0.97, 0.51, 0.47}
\definecolor{darkmagenta}{rgb}{0.55, 0.0, 0.55}
\definecolor{royalblue}{rgb}{0.25, 0.41, 0.88}
\definecolor{ForestGreen}{RGB}{34,139,34}
\definecolor{ao}{rgb}{0.0, 0.5, 1.0}
\definecolor{hippiepink}{rgb}{0.6835, 0.2539, 0.3281}
\definecolor{iy}{rgb}{0.0, 0.36, 0.05}
\definecolor{zb}{RGB}{0, 179, 184}

\definecolor{ra}{rgb}{0.05, 0.5, 0.06}
\definecolor{rb}{rgb}{0.9, 0.17, 0.31}
\definecolor{rc}{rgb}{0.06, 0.2, 0.65}
\definecolor{rd}{rgb}{0.93, 0.53, 0.38}
\definecolor{re}{rgb}{0.1, 0.5, 0.5}

\definecolor{cq}{rgb}{0.0, 0.5, 1.0}

\definecolor{shadecolor}{RGB}{180,180,180}
\definecolor{darkgray}{rgb}{0.86, 0.86, 0.86}
\usepackage[framemethod=tikz]{mdframed}
\newcommand{\fancycommand}[1]{%
\begin{mdframed}[backgroundcolor=darkgray, linecolor=black, linewidth=0.5pt]
    \texttt{#1}%
\end{mdframed}%
}

\newif\ifcamerareadyiterations
\camerareadyiterationsfalse

\newif\ifcameraready
\camerareadytrue
\ifcameraready
  \newcommand\nbcr[2]{#2}
  \newcommand\atbcr[2]{#2}
  \newcommand{\nbcomment}[1]{}
   \newcommand{\atbcrcomment}[1]{}
    
\else 
     \newcommand\nbcr[2]{\ifnum#1=\value{version}\textcolor{red}{#2}\else{\textcolor{blue}{#2}}\fi}
    \newcommand\atbcr[2]{\ifnum#1=\value{version}\textcolor{orange}{#2}\else{\textcolor{blue}{#2}}\fi}
    \newcommand{\atbcrcomment}[1]{\todo[size=\scriptsize, linecolor=orange, bordercolor=orange, backgroundcolor=white]{\textcolor{orange}{\textbf{@atb:} #1}}}

    \usepackage[colorinlistoftodos,prependcaption,textsize=small]{todonotes}
    \usepackage{marginnote} 
    \let\marginpar\marginnote
    \paperwidth=\dimexpr \paperwidth + 4cm\relax
    \oddsidemargin=\dimexpr\oddsidemargin + 2cm\relax
    \evensidemargin=\dimexpr\evensidemargin + 2cm\relax
    \marginparwidth=\dimexpr \marginparwidth + 2cm\relax
    \newcommand{\nbcomment}[1]{\todo[size=\scriptsize, linecolor=nbs, bordercolor=nbs, backgroundcolor=white]{\textcolor{nbs}{\textbf{@nb:} #1}}}
\fi

\newcommand{\versionnum}[0]{1.1}

\newif\ifartifactdraft
\artifactdraftfalse

\ifartifactdraft
    \newcommand{\aedraft}[1]{\textcolor{blue}{#1}}
\else
    \newcommand{\aedraft}[1]{\textcolor{black}{#1}}

\fi

\newif\ifdraft
\draftfalse

\newif\ifarxivdraft
\arxivdraftfalse

\newif\ifrevisiondraft
\revisiondraftfalse

\ifrevisiondraft

    \newcommand{\revcommon}[1]{\textcolor{cq}{#1}}
    \newcommand{\rva}[1]{\textcolor{ra}{#1}}
    \newcommand{\rvb}[1]{\textcolor{black}{#1}}
    
    \newcommand{\rvd}[1]{\textcolor{rd}{#1}}

    \newcommand\revmd[2]{\todo[linecolor=#1,backgroundcolor=#1!15,bordercolor=#1]{\textcolor{black}{\textbf{#2}}}}

    \newcommand{\labela}[1]{\revmd{ra}{A#1}}

    \newcommand{\labeld}[1]{\revmd{rd}{D#1}}

\else
    \newcommand{\revcommon}[1]{{#1}}
    \newcommand{\rva}[1]{{#1}}
    \newcommand{\rvb}[1]{{#1}}
    
    \newcommand{\rvd}[1]{{#1}}

    \newcommand\revmd[2]{}

    \newcommand{\labela}[1]{\revmd{ra}{A#1}}

    \newcommand{\labeld}[1]{\revmd{rd}{D#1}}

\fi

\ifdraft
    \usepackage[colorinlistoftodos,prependcaption,textsize=small]{todonotes}
    \paperwidth=\dimexpr \paperwidth + 4cm\relax
    \oddsidemargin=\dimexpr\oddsidemargin + 2cm\relax
    \evensidemargin=\dimexpr\evensidemargin + 2cm\relax
    \marginparwidth=\dimexpr \marginparwidth + 2cm\relax
    \newcommand{\nb}[1]{\textcolor{nbs}{#1}}
    \newcommand{\atb}[1]{\textcolor{ao}{#1}}
    \newcommand{\agy}[1]{\textcolor{hippiepink}{#1}}
    \newcommand{\iey}[1]{\textcolor{iy}{#1}}
    \newcommand{\zb}[1]{\textcolor{zb}{#1}}

    \newcommand{\nbcomment}[1]{\todo[size=\scriptsize, linecolor=orange, bordercolor=orange, backgroundcolor=white]{\textcolor{nbs}{\textbf{@nb:} #1}}}
    \newcommand{\intexttodo}[1]{\textcolor{tangerine}{TODO: #1}}
    \newcommand{\atbcomment}[1]{\todo[size=\scriptsize, linecolor=ao, bordercolor=ao, backgroundcolor=white]{\textcolor{ao}{\textbf{@atb:} #1}}}
    \newcommand{\agycomment}[1]{\todo[size=\scriptsize, linecolor=ao, bordercolor=ao, backgroundcolor=white]{\textcolor{hippiepink}{\textbf{@gy:} #1}}}
    \newcommand{\zbcomment}[1]{\todo[size=\scriptsize, linecolor=zb, bordercolor=zb, backgroundcolor=white]{\textcolor{hippiepink}{\textbf{@zb:} #1}}}
    \newcommand{\nmgcomment}[1]{\todo[size=\scriptsize, linecolor=forestgreen, bordercolor=forestgreen, backgroundcolor=white]{\textcolor{cornflowerblue}{\textbf{@nmg:} #1}}}

    \newcommand{\mscomment}[1]{\todo[size=\scriptsize, linecolor=blue, bordercolor=red, backgroundcolor=white]{\textcolor{hippiepink}{\textbf{@ms:} #1}}}
    \newcommand{\ieycomment}[1]{\todo[size=\scriptsize, linecolor=orange, bordercolor=orange, backgroundcolor=white]{\textcolor{iy}{\textbf{@iey:} #1}}}

\newcommand\param[1]{{\color{blue}{#1}}}

\else 
    
    \newcommand{\nb}[1]{{#1}}
    \newcommand{\agy}[1]{#1}
    \newcommand{\atb}[1]{#1}
    \newcommand{\iey}[1]{#1}
    \newcommand{\zb}[1]{#1}
    \newcommand{\intexttodo}[1]{}
    \newcommand{\atbcomment}[1]{}
    \newcommand{\agycomment}[1]{}
    \newcommand{\zbcomment}[1]{}
    \newcommand{\mscomment}[1]{}
    \newcommand{\ieycomment}[1]{}
    \newcommand{\nmgcomment}[1]{}
    \newcommand\param[1]{{\color{black}{#1}}}

\fi

    \newcommand\revmid[2]{}
    \newcommand{\revat}[1]{\textcolor{black}{#1}}

    \newcommand{\revbt}[1]{\textcolor{black}{#1}}
    \newcommand{\revb}[1]{\revmid{black}{QB#1}}

    \newcommand{\revct}[1]{\textcolor{black}{#1}}
    \newcommand{\revc}[1]{\revmid{black}{QC#1}}

    \newcommand{\revdt}[1]{\textcolor{black}{#1}}
    \newcommand{\revd}[1]{\revmid{black}{QD#1}}

    \newcommand{\revet}[1]{\textcolor{black}{#1}}
    \newcommand{\reve}[1]{\revmid{black}{QE#1}}

    \newcommand{\cqt}[1]{\textcolor{black}{#1}}

\ifarxivdraft
    \newcommand{\nba}[1]{\textcolor{nbs}{#1}}
    \renewcommand{\intexttodo}[1]{\textcolor{tangerine}{TODO: #1}}
    \renewcommand{\agycomment}[1]{\todo[size=\scriptsize, linecolor=ao, bordercolor=ao, backgroundcolor=white]{\textcolor{hippiepink}{\textbf{@gy:} #1}}}
    \renewcommand{\zbcomment}[1]{\todo[size=\scriptsize, linecolor=zb, bordercolor=zb, backgroundcolor=white]{\textcolor{hippiepink}{\textbf{@zb:} #1}}}
        \paperwidth=\dimexpr \paperwidth + 4cm\relax
    \oddsidemargin=\dimexpr\oddsidemargin + 2cm\relax
    \evensidemargin=\dimexpr\evensidemargin + 2cm\relax
    \marginparwidth=\dimexpr \marginparwidth + 2cm\relax
\else
    \newcommand{\nba}[1]{\textcolor{black}{#1}}
    \renewcommand{\intexttodo}[1]{}
    \renewcommand{\agycomment}[1]{}
    \renewcommand{\zbcomment}[1]{}
\fi

\newcommand{\head}[1]{{\noindent\textbf{#1.}\xspace}} % for heading of a paragraph
\usepackage{tikz}

\newcommand*\circled[1]{\tikz[baseline=(char.base)]{\node[shape=circle,fill,inner sep=0.5pt] (char) {\textcolor{white}{#1}};}}

\newcommand{\attackvone}{IMPACT-PnM}
\newcommand{\attackvtwo}{IMPACT-PuM}

\usepackage{dblfloatfix}
\usepackage{array}
 \usepackage{booktabs}

\newcommand{\dataMovementCitations}[0]{\cite{mutlu2013memory,mutlu2015research,dean2013tail,kanev_isca2015,ferdman2012clearing,wang2014bigdatabench,mutlu2019enabling,mutlu2019processing,mutlu2020intelligent,mutlu2020modern,boroumand2018google,wang2016reducing,pandiyan2014quantifying,koppula2019eden,kang2014co,mckee2004reflections,wilkes2001memory,kim2012case,wulf1995hitting,ahn2015scalable,hsieh2016transparent,wang2020figaro}}

\newcommand{\pimCitations}[0]{\cite{pandiyan2014quantifying,kanev2015profiling,mutlu2019processing,paul2015harmonia,ware2010architecting,lefurgy2003energy,boroumand2018google,vogelsang2010understanding,wulf1995hitting,oliveira2021damov,mutlu2013memory,mutlu2015research}}

\newcommand{\pnmCitations}[0]{\cite{pawlowski2011hybrid,loh2008isca,hmc_spec,nai2017graphpim,kim2018grim,gao2017tetris,kim2016neurocube,ghiasi2022genstore,lee2021hardware,rosenfeld2014performance,hsieh2016transparent,besta2021sisa,singh2019napel,ke2021near,oliveira2022accelerating}}

\newcommand{\pumCitations}[0]{\cite{aga2017compute,flashcosmos,fujiki2019duality,gao2019computedram,li2016pinatubo,seshadri2013rowclone,seshadri2017ambit,song2018graphr,xin2020elp2im,gao2022frac,yuksel2023pulsar,yuksel2024functionally,hajinazar2021simdram,peng2023chopper,shahroodi2023swordfish,deoliveira2024mimdram}}

\newcommand{\secref}[1]{§\ref{#1}}
\newcommand{\figref}[1]{Fig.~\ref{#1}}

\usepackage{listings}
\lstdefinestyle{customc}{
  belowcaptionskip=1\baselineskip,
  breaklines=true,
  frame=single,
  xleftmargin=0.5\parindent,
  language=C,
  showstringspaces=false,
  basicstyle=\footnotesize\ttfamily,
  keywordstyle=\bfseries\color{green!40!black},
  commentstyle=\itshape\color{tangerine},
  identifierstyle=\color{black},
  stringstyle=\color{orange},
  numbers=left,
  stepnumber=1,
  numbersep=-5pt,
  escapeinside={*@}{@*}, % Added escape sequence
  captionpos=b
}

\newif\ifpaper
\papertrue % or
\ifpaper

\else

\fi

\newcommand{\affilETH}{$^\dagger$}
\newcommand{\affilZulal}{$^{\dagger\ddag}$}
\newcommand{\affilFirst}{$^{\dagger\ast}$}

\fancypagestyle{iterationsfirstpage}
{
    \fancyhead{}
    \fancyhead[C]{
    \textcolor{red}{CONFIDENTIAL DRAFT -- DO NOT DISTRIBUTE -- TO APPEAR IN DSN'25} \\ 
    \emph{\textcolor{blue}{Version \versionnum~---~\today, \ampmtime}}
    }
}

\begin{document}

%%%%%%%%%%%---SETME-----%%%%%%%%%%%%%
\title{{Revisiting Main Memory-Based Covert and Side Channel Attacks\\in the Context of Processing-in-Memory}}

\author{
{F.~Nisa~Bostanc\i}\affilFirst\quad
{Konstantinos~Kanellopoulos}\affilFirst\quad
{Ataberk~Olgun}\affilETH\qquad \\
{A.~Giray~Ya\u{g}l{\i}k\c{c}{\i}}\affilETH\qquad 
{İsmail~Emir~Y\"{u}ksel}\affilETH\qquad 
{Nika~Mansouri~Ghiasi}\affilETH\qquad \\
{Z\"{u}lal~Bing\"{o}l}\affilZulal\qquad  
{Mohammad~Sadrosadati}\affilETH\qquad
{Onur~Mutlu}\affilETH\vspace{0mm}\\\\
{$^\dagger${ETH Z{\"u}rich}\quad$^\ddag${Bilkent University}}
}

%%%%%%%%%%%%%%%%%%%%%%%%%%%%%%%%%%%%

\maketitle

%Enables the camera ready header and footer
\ifcamerareadyiterations 
  \thispagestyle{iterationsfirstpage}
  \pagestyle{plain}
  \pagenumbering{arabic}
\else
    \renewcommand{\headrulewidth}{0pt}
    \fancypagestyle{firstpage}{
        \fancyhead{} % clear all header and footer fields
        \fancyhead[C]{
        \begin{tikzpicture}[remember picture,overlay]
        \node [xshift=130mm,yshift=-10mm]
        at (current page.north west) {{\includegraphics[width=2.4cm]{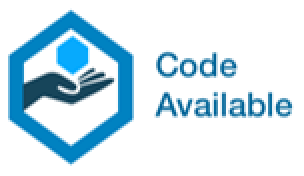}}} ;
        \node [xshift=157mm,yshift=-10mm]
        at (current page.north west) {{\includegraphics[width=2.4cm]{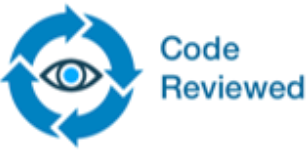}}} ;
        \node [xshift=186mm,yshift=-10mm]
        at (current page.north west) {{\includegraphics[width=2.8cm]{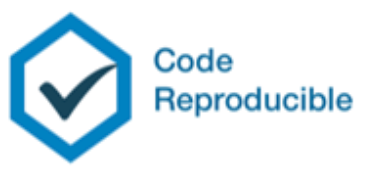}}} ;
        \end{tikzpicture}
         
       % \fontsize{12pt}{12pt}\selectfont\thepage
      } % except the center  
    \renewcommand{\footrulewidth}{0pt}
    }
  \thispagestyle{firstpage}
  % \thispagestyle{empty}
  % \pagestyle{empty}
  % \pagenumbering{gobble}

\fi

\pagenumbering{arabic}

%%%%%% -- PAPER CONTENT STARTS-- %%%%%%%%
\newcounter{version}
\setcounter{version}{3}
\begin{abstract}
We introduce \X{}, a {set of} high-throughput main memory-based {timing} {attacks that leverage characteristics of \nbcr{0}{\textit{processing-in-memory} (PiM)} architectures to establish covert and side channels.}
\X{} 
enables high-throughput communication and {private information leakage} {by exploiting the shared DRAM row buffer}. {To achieve high throughput,
\X{} (i)} eliminates expensive cache bypassing steps required by processor-centric memory-based {timing} attacks and (ii) leverages the intrinsic parallelism of PiM operations. {We showcase two applications of \X{}.} 
{First, we build two covert channels} that leverage {different PiM approaches (i.e., \textit{processing-near-memory} and \textit{processing-using-memory})} to establish high-throughput covert \nbcr{3}{communication} channels. \nbcr{0}{Our covert channels} achieve \param{8.2 Mb/s} and \param{14.8 Mb/s} communication throughput, respectively, which is \param{3.6$\times$} and \param{6.5$\times$} \nbcr{0}{higher} than the state-of-the-art main memory-based covert channel.
\nba{Second, we showcase a side-channel attack that leaks private information of concurrently-running victim applications} \nbcr{0}{with a low error rate.}
\nbcr{0}{Our source-code is openly and freely available at 
\texttt{https://github.com/CMU-SAFARI/IMPACT}}.
\end{abstract}

\renewcommand{\thefootnote}{\fnsymbol{footnote}}
\footnotetext[1]{F. N. Bostanc{\i} and K. Kanellopoulos are co-primary authors.}%
\renewcommand{\thefootnote}{\arabic{footnote}}
\setstretch{0.99}
\section{Introduction}
\label{sec:intro}

\nbcr{0}{\emph{Data movement} between computation units (e.g., CPUs, GPUs, ASICs) and main memory (e.g., DRAM) is a major \emph{performance} and \emph{energy bottleneck} in current processor-centric computing systems~\dataMovementCitations{}, and is expected to worsen due to the increasing data intensiveness of modern applications, e.g., machine learning~\cite{brown2020language,devlin2019bert,boroumand2021google,oliveira2022accelerating,heo2024neupims,zhou2022transpim,park2024attacc,seo2024ianus, rhyner2024pim,yun2024duplex,jang2024smart} and genomics~\cite{alser2020accelerating,singh2021fpga,alser2022from,kim2018grim,ghiasi2022genstore,ghiasi2024megis,cali2020genasm,cali2022segram,cali2017nanopore}.
To mitigate the overheads caused by data movement, various works propose \gls{pim} architectures~\pimCitations{}.
There are two main approaches to \gls{pim}~\cite{ghose2019processing, mutlu2020modern, mutlu2019processing}:
(i)~\gls{pnm}~\pnmCitations{}, where computation logic is added near the memory arrays (e.g., in a DRAM chip, next to each bank, or at the logic layer of a 3D-stacked memory~\cite{HMC2, lee20151, lee2016simultaneous, jedec2021hbm, hbm3, kim2024present, ahn2015scalable,PEI,hsieh2016transparent,boroumand2018google,boroumand2022polynesia,boroumand2021google,boroumand2019conda,singh2021fpga,oliveira2022accelerating}); and (ii)~\gls{pum}~\pumCitations{}, where computation is performed by exploiting the analog operational properties of the memory circuitry.}

\nbcr{0}{Prior works from industry and academia demonstrate the \nbcr{2}{large} performance and energy benefits of \gls{pim} architectures. A set of \gls{pim} techniques are already implemented in real products~\cite{devaux2019true,ke2021near,lee20221ynm,niu2022184qps,kwon202125} and many more are expected to be adopted in the near future.}
Therefore, it is important to investigate the security \nbcr{2}{aspects} of such techniques and implement countermeasures to avoid potential widespread security vulnerabilities that their deployment can cause.
\nba{Unfortunately, {\emph{no}} prior work analyzes and evaluates the security of emerging PiM architectures {against timing covert- and side-channel attacks}}.

\nbcr{0}{In this work, we analyze \gls{pim} architectures and show that the adoption of \gls{pim} architectures creates opportunities for critical main memory-based timing attacks due to two reasons. First, to eliminate data movement, \gls{pim} architectures provide direct access to main memory, which is a key building block for high-throughput main memory-based timing attacks. Second, defenses against these attacks either incur high performance overheads or are \emph{not} applicable to \gls{pim} architectures.}

\noindent
{\textbf{1. Direct Access to Main Memory.}} 
\nbcr{0}{Main memory-based timing attacks establish covert communication channels or leak secrets by observing and changing the shared main memory states (e.g., DRAM row buffer states). To achieve high throughput,} \nba{{timing} attacks {\nbcr{0}{require fast and reliable} access \nbcr{0}{to} main memory.}
In today's systems, {it is difficult to access main memory with high frequency due to the deep cache hierarchies that filter the memory requests \nbcr{0}{and incur additional latency overheads\nbcr{1}{~\cite{bera2022hermes}}.}}\footnote{We provide analysis as to why prior attack primitives that bypass caches do \emph{not} provide \textit{fast} and \textit{reliable} \nbcr{1}{enough} direct main memory access in \secref{sec:motivation_establishing}.} 
}

{The adoption of \gls{pim} techniques changes how applications and computation units access {main memory}: (i) due to area and thermal dissipation limitations, compute units in \gls{pnm} architectures have small or no caches that allow the userspace applications to bypass caches more easily (i.e., with fewer \nbcr{1}{instructions and \nbcr{3}{lower} latency}), and (ii) \gls{pum} architectures provide the ability to offload \nbcr{3}{to main memory} specialized PiM instructions (e.g., bulk data copy and initialization~\cite{seshadri2013rowclone,chang2016low,gao2019computedram,olgun2022pidram,deoliveira2024mimdram}, and bitwise/arithmetic operations~\cite{seshadri2017ambit,seshadri2019dram,gao2019computedram,gao2022frac,xin2020elp2im,besta2021sisa,deoliveira2024mimdram,yuksel2024functionally,yuksel2023pulsar,hajinazar2021simdram}) that completely bypass the deep cache hierarchies of modern processors. \nba{As a result, PiM architectures provide userspace applications with \textit{fast} and \textit{reliable} direct main memory access that can be exploited by attackers for high-throughput timing covert- and side-channel attacks.}}

\noindent
{\textbf{2. \atb{Hard-to-Mitigate} with Practical Defenses.}}
{A fundamental solution to memory-based timing attacks is to eliminate the timing channel \nbcr{0}{completely (e.g., by enforcing constant-time memory accesses or introducing intrusive \nbcr{1}{and restrictive} memory partitioning between different security contexts)}. However, eliminating the timing channel incurs \nbcr{1}{very} high performance overheads (as shown in Section~\ref{sec:mitigations}). 
Therefore, \agy{several prior works~\cite{yan2017secure,mushtaq2018nights,alam2017performance,chiappetta2016real}} \nbcr{0}{propose} defenses and detection mechanisms that restrict \nbcr{1}{the use of} timing channels with cache management-based methods,
such as restricting the \nbcr{2}{userspace applications'} access to cache management instructions (e.g., \texttt{clflush} instruction in x86 systems~\cite{yan2017secure}) and detecting timing attacks based on {cache access patterns (e.g., a} high number of cache misses {can indicate a timing attack}){~\cite{mushtaq2018nights,alam2017performance,chiappetta2016real}}. These \nbcr{0}{defenses} are \nbcr{1}{\emph{inapplicable}} to PiM \nbcr{0}{architectures because these architectures provide \nbcr{1}{\emph{direct}} access to memory by bypassing} the cache hierarchy.}

\agy{Based on our analysis}, 
we introduce \underline{IMPA}C\underline{T}, a \nba{set of high-throughput \underline{I}n-\underline{M}emory \underline{P}rocessing-based {timing} \underline{At}tack\nba{s} that leverage direct and fast main memory accesses enabled by PiM architectures.} 
Using direct access \nbcr{1}{to DRAM} via PiM operations, \X{} \nba{observes and exploits the shared DRAM row buffer \nbcr{0}{states}. We demonstrate that \X{}} 
achieves high communication \nba{and information leakage} throughput 
by (i) \nba{eliminating} expensive cache bypassing steps required by other main memory- and cache-based timing covert- and side-channel attacks, and (ii) leveraging the intrinsic parallelism of PiM operations. 
\nba{We showcase this finding on two covert- and one side-channel attacks. To protect future systems against \X{}, we \nbcr{0}{\nbcr{1}{describe} and analyze four} defense mechanisms.}

\noindent
\textbf{\attackvone{} Covert Channel Variant.}
We build a covert channel \nbcr{0}{exploiting a \gls{pnm} architecture \nbcr{3}{called PiM-enabled instructions~\cite{PEI}} that enables executing instructions in compute units near DRAM}. \nbcr{0}{In this variant, the receiver uses \gls{pim}-enabled instructions to measure the time it takes to execute a simple arithmetic operation on a set of DRAM rows and detect row buffer conflicts. The sender transmits a message by encoding the message with row buffer conflicts in selected addresses and carefully creating conflicts with the receiver's accesses using \gls{pim}-enabled instructions.} 
\nbcr{0}{By executing instructions near memory,}
\attackvone{} achieves 8.2 Mb/s communication throughput, which is \param{3.6$\times$} faster than the state-of-the-art main memory-based \nbcr{1}{covert-channel} attack~\cite{pessl2016drama}. 

\noindent
\textbf{\attackvtwo{} Covert Channel Variant.}
\nbcr{0}{We build a covert channel exploiting a \gls{pum} technique \nbcr{3}{called RowClone~\cite{seshadri2013rowclone}} that enables bulk data copy and initialization. In this variant, the sender transmits a message \nbcr{1}{with \nbcr{3}{in-memory} row-copy operations} \nbcr{3}{by copying rows in different DRAM banks in parallel} for different bits of the message. The receiver decodes the message by issuing \nbcr{1}{row-copy} operations and measuring their latency. By exploiting the parallelism in memory,} 
\attackvtwo{} achieves a communication throughput of \param{14.8 Mb/s}, which is \param{6.5$\times$} faster than the state-of-the-art~\cite{pessl2016drama}. 

\noindent
{\textbf{\X{} Side-Channel Attack.} \nba{We build a PnM-based \X{} side-channel attack that exploits \nbcr{0}{\gls{pim}-enabled instructions}\nbcr{1}{~\cite{PEI}} to leak private information \nbcr{0}{of} a concurrently-running application.}
We showcase \nba{this} side-channel attack on \nbcr{1}{genomic} \zb{read mapping, one of the fundamental applications in DNA sequence analysis where preserving the privacy of the human genomic data is crucial~\cite{alser2020accelerating,alser2021technology,alser2022from,lin2004genomic,lu2021methods,wang2009learning,arshad2021analysis}.} Our side-channel attack leverages \nbcr{0}{\gls{pim}-enabled instructions~\cite{PEI} to leak} the private characteristics of a user's \nbcr{1}{query} genome by observing the memory access patterns of the read mapping application. Our evaluation demonstrates that our side-channel attack leaks the properties of a \nbcr{1}{query} genome at a throughput of 7.6 Mb/s with 96\% accuracy.}

\noindent
\textbf{Defense Mechanisms.}
Based on our security analysis, we discuss and evaluate potential defense mechanisms to mitigate \X{} attacks \nb{by eliminating the timing channel \nbcr{0}{and reducing the attack throughput}. We} \nbcr{1}{describe} the performance versus security trade-offs of \nb{these} defenses. \nbcr{0}{We eventually conclude that mitigating \X{} incurs high performance overheads} \nbcr{1}{and more research is needed to find low-overhead solutions.}

{{This paper makes} the following contributions:}
\begin{itemize}
    \item \nba{To our knowledge, this is the first work that analyzes and evaluates the security of emerging PiM architectures against timing covert-\nbcr{1}{channel} and side-channel attacks.} 
    \item \nba{We introduce \X{}, a new set of high-throughput main memory-based timing attacks that leverage PiM operations to gain fast and reliable access to main memory.}
    \item \nba{We \nbcr{1}{demonstrate} two \X{} covert channel {variants}, \attackvone{} and \attackvtwo{} that exploit PnM and PuM techniques, respectively. \attackvone{} allows the sender and the receiver to exchange messages using PiM-enabled instructions. {\attackvtwo{}} further leverages the parallelism provided by \nbcr{3}{bulk in-memory copy operations like} \nbcr{1}{RowClone}. We show that both attacks outperform the state-of-the-art main-memory-based covert-channel attacks in terms of communication throughput.}
    \item {We \nbcr{1}{demonstrate} an \X{} side-channel attack on \nbcr{2}{genomic} read mapping by leveraging PnM, \nbcr{3}{using \gls{pim}-enabled} instructions. We show that our side-channel attack leaks sensitive information with high throughput and a low error rate.}
    \item We \nbcr{1}{\nbcr{2}{describe} and evaluate} \agy{the trade-offs of four} defense mechanisms to protect \agy{PiM-enabled} systems against \X{}'s \agy{covert- and side-channel attacks}. \nbcr{1}{We call for future research to find low-overhead solutions.} 
\end{itemize}

\glsresetall
\section{Background}
\nbcr{1}{This section provides a concise overview of 1)~DRAM organization and operation, 2)~\nbcr{2}{processing}-in-memory, and 3)~\nbcr{2}{main} memory-based timing attacks.}

\subsection{DRAM Organization and Operation}

{\textbf{Organization.} 
\figref{fig:dram_org} shows the hierarchical organization of a modern DRAM-based main memory. \nbcr{1}{A memory channel connects the processor (CPU) to a DRAM module, where a module consists of multiple DRAM ranks.}
A rank consists of multiple DRAM chips that operate in lockstep. 
Each DRAM chip contains multiple DRAM banks that can be accessed independently. A DRAM bank is organized as a two-dimensional array of DRAM cells, where a row of cells is called a DRAM row. A DRAM cell consists of 1) a storage capacitor, which stores one bit of information in the form of electrical charge, and 2) an access transistor, which connects the capacitor to \textit{the row buffer} through a bitline controlled by a wordline.

\begin{figure}[ht]
    \centering
    \includegraphics[width=0.9\linewidth]{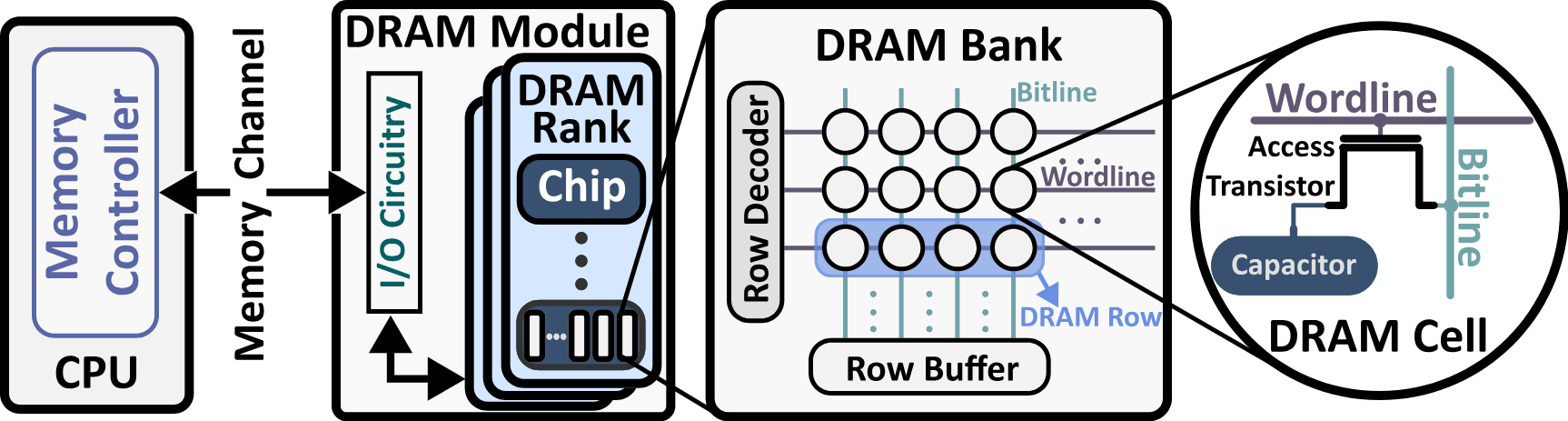}
    \caption{Organization of a typical modern DRAM module.}
    \label{fig:dram_org}
\end{figure}

\noindent{\textbf{Operation.} 
{To perform a read or write operation, the memory controller first needs to open a row, i.e., copy the data of the cells in the row to the \textit{row buffer}. To open a row, the memory controller issues an activation command to a bank by specifying the row address to open. After activation is complete, the memory controller issues either read \nbcr{2}{or} write commands to read or write a DRAM word within the activated row. Subsequent accesses to the same row can be served quickly without fetching the data to the row buffer, creating a \emph{row hit}~\cite{keeth2001dram}. To access data from another DRAM row in the same bank, the memory controller must first close the currently open row by issuing a bank precharge command, creating a \emph{row conflict/miss}.}
}

\subsection{Processing-in-Memory}
\gls{pim} moves computation closer to where the data resides, alleviating the data movement bottleneck between memory and the processor~\pimCitations{}. There are two main PiM approaches~\cite{ghose2019processing, mutlu2020modern, mutlu2019processing}: 1) \gls{pnm} and 2) \gls{pum}.
{\gls{pnm} \iey{places \nbcr{3}{computation} logic near memory arrays (e.g., \nbcr{2}{in a DRAM chip, next to each bank, or at the logic layer} of 3D-stacked memory~\pnmCitations{}).} The \gls{pnm} logic can execute application regions (e.g., instructions~\cite{PEI,nai2017graphpim}, functions~\cite{oliveira2021damov}, application threads~\cite{ahn2015scalable,song2018graphr}) depending on the design. PnM computation units typically employ small caches due to 1) area constraints and 2) the ineffectiveness of caches \nbcr{2}{in} improving the performance of a wide range of memory-intensive applications. Therefore, various
systems~\cite{nai2017graphpim,PEI,oliveira2021damov} \agy{offload} the memory-intensive application \agy{regions} on PnM units, while executing more cache-friendly \agy{application regions} on the host CPU with a deeper cache hierarchy.
PuM \iey{uses analog operational properties of the memory circuitry (e.g., DRAM) to enable massively parallel in-memory computation}~\pumCitations{}.

\subsection{Main Memory Timing Attacks}
\label{sec:mainmemorytiming}

Microarchitectural timing attacks exploit the observable microarchitectural state changes to propagate messages between processes and leak secrets. These attacks measure the time to complete an operation that depends on a shared system structure (e.g., accessing a memory location by going through the shared caches) to infer the state of the shared structure.  

Main memory timing attacks exploit several shared structures in the main memory. One such structure is the DRAM row buffer that acts as an internal cache inside DRAM banks. An attacker can understand \nbcr{1}{whether} a row is recently accessed (i.e., \textit{already in the row buffer}) by measuring  \nb{the time it takes to access} that row. Using this observation, DRAMA~\cite{pessl2016drama} builds a side-channel attack to leak keystrokes in a real system. Several prior works \nbcr{2}{(e.g.,~\cite{pessl2016drama,xiao2016one, kwong2020rambleed,moscibroda2007memory,bhattacharya2016curious})} leak DRAM address mapping functions. 
Other \agy{main memory-based} attacks exploit the memory bus \nbcr{2}{contention}~\cite{zhenyu2012whispers} and memory deduplication~\cite{xiao2013security,lindemann2018memory} and (de)compression~\cite{kelsey2002compression,rizzo2012crime,gluck2013breach,be2013perfect,vanhoef2016heist,van2016request,karakostas2016practical,tsai2020safecracker,schwarzl2021practical} latencies.}

\section{A Case for PiM-based {Timing} Attacks}

The adoption of \gls{pim} architectures has been gaining momentum in recent years {because they provide high performance and low energy consumption by \nba{reducing data movement}.}
{Therefore, it is timely and critical to analyze and evaluate the security of PiM architectures to prevent any vulnerabilities from causing widespread susceptibility as PiM architectures become widely available. 
\nba{Unfortunately, {\emph{no}} prior work analyzes and evaluates the security of emerging PiM architectures {against timing covert- and side-channel attacks}}.

After a thorough analysis of existing PiM architectures, we show that \nbcr{1}{adopting} PiM solutions creates opportunities for high-throughput main memory-based timing attacks for two main reasons. 
{First, the adoption of PiM solutions provides userspace applications with fast and reliable direct access to main memory \nbcr{1}{to reduce data movement}.}
Second, 
\nbcr{1}{defenses against these attacks either incur high performance overheads or are \emph{not} applicable to \gls{pim} architectures.}
\nbcr{1}{In this work, we study the impact of \gls{pim} architectures on row buffer-based timing channels and their throughput.}

\subsection{Row Buffer Timing Channel}
A DRAM row buffer acts as a direct-mapped cache inside a DRAM bank and is shared across applications that can access main memory. Accessing a row that is already in the row buffer results in a significantly lower latency because the memory controller does \emph{not} need to issue additional commands \nbcr{1}{(i.e., precharge the bank and fetch the target row).} Therefore,
a \nbcr{1}{userspace} application \agy{can} infer \agy{whether} a row has been accessed recently by measuring the time it takes to access the row.

\nbcr{1}{To demonstrate the latency difference between a row buffer hit and a row buffer conflict in a \gls{pim}-enabled system, we simulate such a system (described in detail in~\secref{sec:method}). We run a microbenchmark that 1) creates memory accesses that result in row buffer hits and conflicts, and 2) measures the memory access latency for each memory request.}
\agy{We observe that a row buffer conflict takes 74 CPU cycles more than a hit \nbcr{2}{for a CPU operating at 2.6 GHz},}
\agy{which} is large enough to detect, \nbcr{1}{which aligns with the observations of prior works~\cite{pessl2016drama,moscibroda2007memory,bhattacharya2016curious}}. \nbcr{1}{Based on this observation, we conclude that} \agy{a \nbcr{1}{userspace} application can exploit the row buffer as a timing channel \nbcr{1}{in a \gls{pim}-enabled system}.}

\subsection{Existing Main Memory Attack Primitives}
\label{sec:motivation_establishing}
Main memory timing attacks are powerful as they can be used to gather information on \textit{all} applications that share main memory. However, these attacks \zb{have limited throughput}
in traditional architectures \nbcr{1}{because accessing main memory \nbcr{2}{\emph{directly}} is difficult due to deep cache hierarchies.}

There are \nbcr{1}{four main attack primitives that enables attackers to} bypass the cache hierarchy \nb{and directly access main memory}: 1) using specialized instructions to flush a \nbcr{1}{target} cache line (e.g., \texttt{clflush}~\cite{intelmanual2016}) \nbcr{1}{\nbcr{2}{from} the cache deterministically}, 2) using cache eviction sets~\cite{liu2015last} \nbcr{1}{to replace a target cache line in the cache} (i.e., by creating many conflicting memory requests to replace the cache line with newly-fetched cache lines), 3) \nbcr{1}{using a (remote) direct memory access engine (i.e., (R)DMA engine)~\cite{arm2000dma,microchip2008direct},} and 4) \nbcr{1}{using} non-temporal memory hints (e.g., \texttt{movnti} \nbcr{1}{instruction}~\cite{intelmanual2016}).

\nbcr{1}{We analyze the attack primitives in terms of efficiency and effectiveness. To do so,}
\nb{we identify four properties of an efficient (i.e., low-latency) and effective attack primitive: 1) avoiding cache lookup overhead, 2) avoiding high-latency memory accesses, 3) creating an easily detectable timing difference, and 4) reliable functionality (e.g., guaranteed to work by the ISA).} Table~\ref{tab:primitives} shows a comprehensive comparison of these \agy{four} attack primitives \nbcr{1}{and \gls{pim} operations} \nb{in terms of these properties}.

\begin{table}[ht]
\centering
\caption{Efficiency and Effectiveness of Attack Primitives}
\label{tab:primitives}
\resizebox{\linewidth}{!}{
\begin{tabular}{c||c|c||c|c|}
\multicolumn{1}{l||}{}                                                                 & \multicolumn{2}{c||}{\textbf{\textit{Low Latency}}}                                                                                                                  & \multicolumn{2}{c|}{\textbf{\textit{Effectiveness}}}                                                                                                                       \\ 
\cline{2-5}
& \begin{tabular}[c]{@{}c@{}}\textbf{No Cache }\\\textbf{ Lookup}\end{tabular} & \begin{tabular}[c]{@{}c@{}}\textbf{No Excessive }\\\textbf{ Memory Accesses}\end{tabular} & \begin{tabular}[c]{@{}c@{}}\textbf{Timing Difference}\\\textbf{ Detectability}\end{tabular} & \begin{tabular}[c]{@{}c@{}}\textbf{ISA }\\\textbf{ Guarantees}\end{tabular}  \\ 
\hhline{=::====|}
\begin{tabular}[c]{@{}c@{}}\textbf{Specialized }\\\textbf{ Instructions}\end{tabular}  & \xmark                                                                       & \cmark                                                                              & \cmark                                                                                      & \cmark                                                                       \\ 
\hline
\textbf{Eviction Sets}                                                                 & \xmark                                                                       & \xmark                                                                              & \cmark                                                                                      & \xmark                                                                       \\ 
\hline
\textbf{DMA/\nbcr{2}{RDMA}}                                                                     & \cmark                                                                       & \cmark                                                                              & \xmark                                                                                      & N/A                                                                          \\ 
\hline
\begin{tabular}[c]{@{}c@{}}\textbf{Non-temporal }\\\textbf{ Memory Hints}\end{tabular} & \xmark                                                                       & \cmark                                                                              & \cmark                                                                                      & \xmark                                                                       \\ 
\hline
\textbf{PiM Operations}                                                                & \cmark                                                                       & \cmark                                                                              & \cmark                                                                                      & \cmark                                                                       \\
\hhline{=:b:====|}
\end{tabular}
}
\end{table}

First, \nb{\textbf{specialized cache flush instructions} have two main disadvantages that \nbcr{1}{limit their} effectiveness \nbcr{1}{\nbcr{3}{at} directly accessing main memory}: 1) they put the write-back (to the main memory) latency of a cache line on the critical path to access a memory location~\cite{intelmanual2016}, and 2) in some modern systems, these instructions are privileged and cannot be used by userspace applications \nbcr{2}{(e.g., ARM processors~\cite{green2017autolock}})} \nbcr{1}{(not shown in the table)}.} 
Second, \nbcr{1}{to directly access main memory}\nbcr{2}{,} \textbf{\nbcr{3}{cache} eviction sets} require many load requests \nbcr{2}{proportionally} to the number of ways in the caches, which \nbcr{2}{increases} with modern designs. \nb{Therefore, the attacks that use them have limited throughput.} \revbt{\nbcr{1}{Even though the} instructions used to create the eviction sets are part of the ISA, the eviction set is not \textit{guaranteed} to evict the target cache line due to the cache replacement policy, cache prefetchers, and other \nbcr{2}{memory system} design choices that are not transparent to the attacker.}
Third, \textbf{DMA and RDMA engines}~\cite{arm2000dma,microchip2008direct} \nb{incur high overheads due to deep software stacks. As a result, attacks using this primitive need to utilize states with larger timing differences to be distinguishable.} An example \nbcr{2}{work} using this primitive~\cite{ustiugov2020bankrupt} showcases a covert-channel attack based on memory contention. However, \nbcr{1}{memory contention-based timing differences} is not as fine-grained as row buffer-based timing differences. Therefore, the communication throughput of these covert channels is limited (in the order of Kb/s). 
\nb{Fourth, the implementation of \textbf{non-temporal memory hints} varies across systems, as system designers are not bound \nbcr{1}{by the instruction set architecture}. \nb{These instructions may prefetch data from the last level cache (\texttt{prefetchnta}), or can be cached in intermediate buffers (\nbcr{1}{e.g., \texttt{movnti} instruction})~\cite{jattke2024zenhammer}.} Consequently, the ISA does \textit{not} guarantee \nbcr{1}{that non-temporal hints will consistently and \textit{directly} access main memory}.}
%and these hints \nb{often require to be} used with memory fences (e.g., \texttt{sfence} and \texttt{mfence})~\cite{intelmanual2016}, which reduces the potential communication throughput of hint-based covert channel attacks. 
In contrast, \textbf{PiM operations} provide fast and reliable direct memory access to userspace applications, which can be exploited for high-throughput covert- and side-channel attacks, \nbcr{2}{as we show in this paper.}

{\subsection{High Throughput Main Memory-based Covert Channels}}
To demonstrate the impact of the deep cache hierarchy latencies {(i.e., cache access and eviction latencies)} on \agy{an attack's} throughput, we simulate a \nbcr{1}{\gls{pim}-enabled} system (described in \secref{sec:method}), establishing a state-of-the-art \nbcr{1}{row buffer-based} covert channel~\cite{pessl2016drama}. 

\subsubsection{\rvb{Attack Scenarios}}
We simulate two attacks. 
First, we evaluate a \emph{baseline \nbcr{2}{main memory} attack} that relies on cache bypassing using cache eviction sets. 
To transmit \nbcr{1}{messages} through the row buffer, \nbcr{1}{the sender and receiver processes} 1) evict the corresponding cache line \nbcr{1}{to directly access main memory} and 2) fetch the target row into the row buffer using an additional memory request. The baseline attack \nbcr{1}{evicts a cache line} by issuing \texttt{N} memory requests to the target cache set in an \texttt{N}-way cache. The actual eviction latency \nbcr{1}{in a real system} can be much higher due to a larger number of memory accesses required to bypass more complex replacement policies employed in today's systems~\cite{walters2025enterprise,qureshi2007adaptive} or due to address translation overheads~\cite{kanellopoulos2023victima,skarlatos2020elastic,kanellopoulos2023utopia,zhaoISCA2022,kanellopoulos2025virtuoso,midgard,karakostas_characterIISWC,karakostas2015}.
Second, we evaluate a \emph{direct memory access attack} that issues only one memory request to fetch the data to the row buffer \agy{\emph{without}} any cache access or evictions.

For both attacks, we measure the time it takes to send a bit through the row buffer, including cache access and eviction latencies, encoding logic-1 values as row buffer hits and logic-0 as row buffer conflicts. We calculate the communication throughput based on the average time to propagate bits from the sender to the receiver. We analyze the impact of (i) the LLC size and (ii) the number of LLC ways on communication throughput to show the limitations in current and future systems with different cache designs. 

\subsubsection{\rvb{Impact of LLC Size on Communication Throughput}}
\figref{fig:throughput-size} shows the impact of the LLC size on (i) the throughput of the \emph{baseline} and \emph{direct memory access} attacks \nbcr{1}{on the primary y-axis \nbcr{2}{(left)}} and (ii) the latency of a cache line eviction \nbcr{1}{on the secondary y-axis \nbcr{2}{(right)}}. The x-axis shows the cache configuration with increasing last-level cache (LLC) sizes and a fixed number of ways (i.e., 16). \nbcr{1}{We} calculate the cache access latency with increasing LLC sizes \nbcr{2}{using CACTI 6.0~\cite{cacti}}. 

\begin{figure}[h]
\centering
\includegraphics[width=\linewidth]{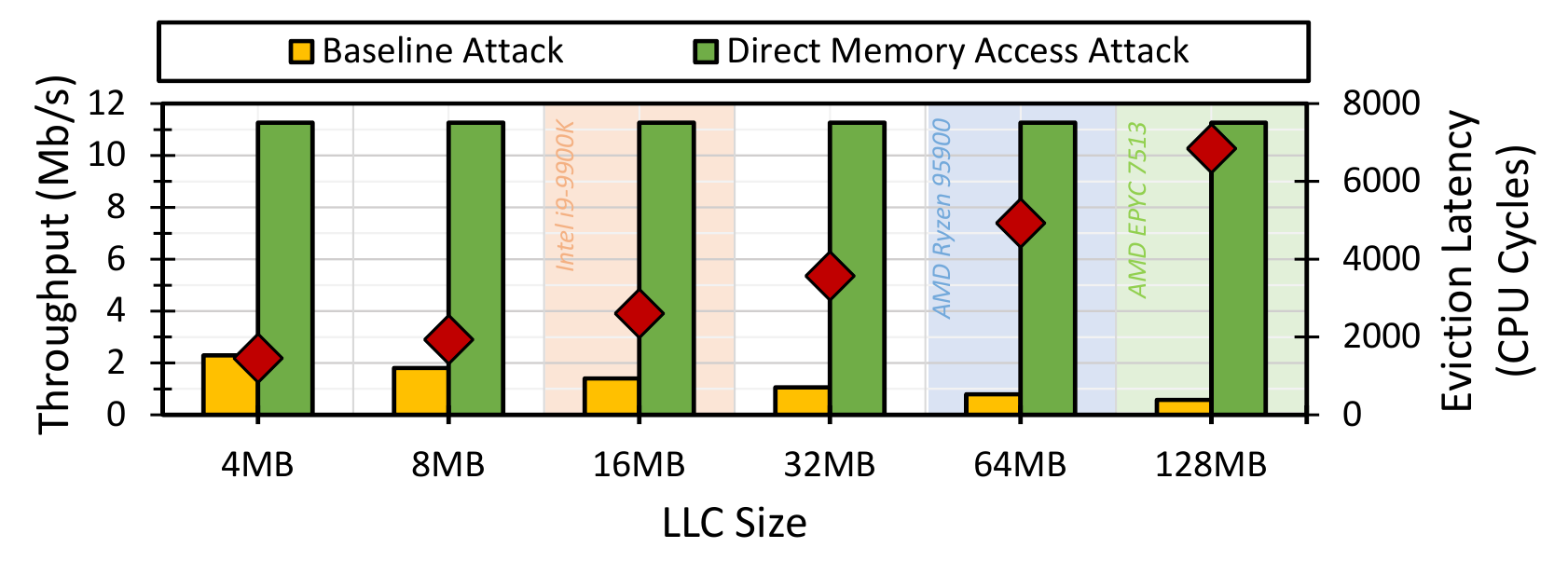}
\caption{Impact of the LLC size on data leakage throughput of covert channels and \nbcr{2}{LLC} eviction latency.}
\label{fig:throughput-size}
\end{figure}

{We make two key observations. First, the communication throughput of the direct memory access attack is \param{$ 11.27$ Mb/s} across all LLC sizes, which is significantly greater than the throughput of up to \param{$ 2.29$ Mb/s}  achieved by the baseline \nbcr{2}{cache eviction sets based} attack.} 
{Second, the baseline attack's throughput decreases with increasing LLC sizes since the cache access and eviction become progressively more expensive. With increasing LLC size, the cache access latency increases, \nbcr{2}{amplifying} the \nbcr{2}{latency} of eviction, which requires multiple cache accesses.}
Therefore, we conclude that \nbcr{1}{with increasing LLC sizes, cache access and eviction latencies increase}, hindering the \agy{baseline attack's} communication throughput, compared to the direct memory access attack, which is independent of the LLC size.

\subsubsection{\rvb{Impact of LLC Associativity on Communication Throughput}}
{\figref{fig:throughput-ways} shows the impact of LLC associativity on (i) the communication throughput of \emph{baseline} and \emph{direct memory access} attacks \nbcr{1}{on the primary y-axis \nbcr{2}{(left)}} and (ii) the eviction latency of a cache line \nbcr{1}{on the secondary y-axis} \nbcr{2}{(right)}. The x-axis shows the increasing number of LLC ways (from 2 to 128) \nbcr{1}{with a fixed total LLC size (e.g., 16 MB).}
We observe that \nbcr{1}{as the number of LLC ways increases, the number of memory requests required to evict a cache line increases, \nbcr{2}{greatly reducing} the baseline attack's communication throughput.}

\begin{figure}[h]
\centering
\includegraphics[width=\linewidth]{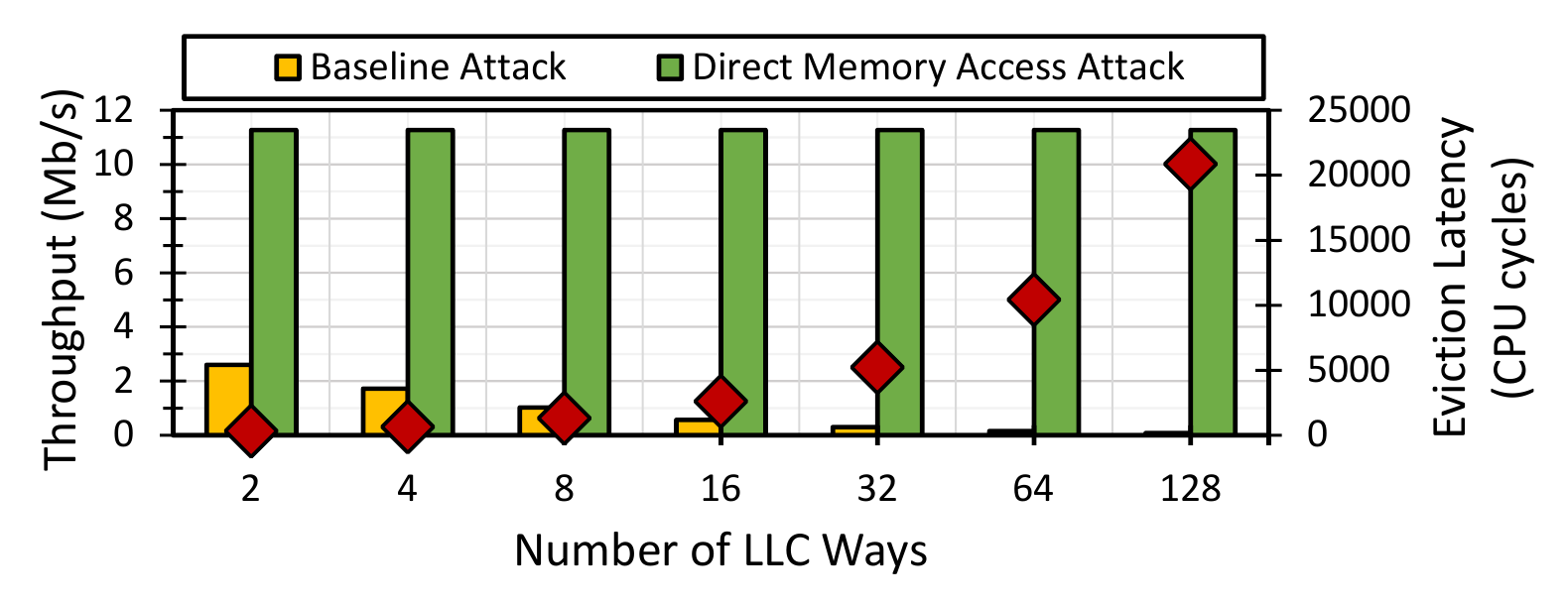}
\caption{Impact of LLC \nbcr{2}{associativity} on the data leakage throughput of covert channels and the eviction latency.}
\label{fig:throughput-ways}
\end{figure}

\nbcr{1}{Based on these two experiments,} we conclude that attacks that communicate with main memory through caches are greatly limited in throughput due to expensive cache access and eviction latencies, which worsen with increasing size and associativity. \nbcr{3}{In contrast}, a covert channel based on direct memory access can achieve high throughput \nbcr{2}{independently} of \nbcr{2}{the} cache configuration, as it does \textit{not} require bypassing the cache hierarchy.

\section{\X{}: A Set of High-Throughput Main Memory Timing Attacks}
\nbcr{3}{We introduce} \X{}, a set of high-throughput main memory-based {timing} attacks that take advantage of the direct memory access {enabled} by emerging \gls{pim} \nbcr{1}{architectures}. 
\X{} achieves high communication \nba{and information leakage} throughput by (i) eliminating expensive cache bypassing steps required by other main memory- and cache-based timing covert- and side-channel attacks and (ii) leveraging the intrinsic parallelism of \gls{pim} operations. 
In this work, we present two covert-channel and one side-channel attacks enabled by \gls{pnm} and \gls{pum} solutions. 

\subsection{PnM-based \X{} Covert-Channel Attack}
\label{sec:pnmattack}

\sloppy
\textbf{Baseline \gls{pnm} Architecture.} 
\gls{pim}-Enabled Instructions (PEI) ~\cite{PEI} is a \gls{pnm} architecture that expands the instruction set architecture 
with \nbcr{3}{instructions} \nbcr{2}{that can be executed in compute units in the processor core or near memory \nbcr{3}{based on a dynamic decision made considering the locality of the required memory accesses}.} 
We choose PEI as the baseline PnM architecture as it introduces a simple and effective model for PnM, and our attack can be generalized for other PnM architectures with similar design components (e.g., FIMDRAM~\cite{kwon202125}).

The baseline PEI architecture has two key components. First, it integrates a PEI Computation Unit (PCU) near each DRAM bank and inside the CPU. 
In-memory PCUs are shared among all cores of the host processor and are responsible for executing PEIs near the respective DRAM bank.
Second, \nbcr{3}{there is} a PEI Management Unit (PMU) that monitors data locality in application regions \nbcr{2}{with a 
\emph{locality monitor,}} and decides where to map each region (either to host-side or memory-side PCUs). This way, \nbcr{3}{the} PEI architecture can execute \nbcr{2}{application} regions with high data locality on the host side to benefit from \nbcr{2}{the cache hierarchy}. \nbcr{3}{In contrast, if an application region does not benefit from the cache hierarchy, it is executed on the memory-side PCU to benefit from the lower memory access latency.}

\textbf{Attack Overview.}
\nba{Before the attack, the sender and the receiver co-locate their data in the same \nbcr{3}{set of} DRAM banks. To achieve this, one process uses memory massaging techniques to place its data \nbcr{3}{in a page located} in the same \nbcr{3}{DRAM} bank as the other process, as used in many prior works \nbcr{3}{(e.g.,~\cite{kwong2020rambleed,pessl2016drama,van2016drammer,vandramaqueen})}.}
\nbcr{2}{To incentivize the \gls{pnm} architecture to execute the instruction in a memory-side PCU \nbcr{3}{and directly access DRAM}, the attack processes access different cache lines in the allocated rows, and allocate and use multiple rows in each bank.}

\attackvone{} uses PEI operations and exploits the latency difference between a DRAM row buffer hit and a conflict in a shared DRAM bank to exchange messages between \nb{\agy{the}} sender and \nb{\agy{the}} receiver. 
The sender transmits bits of the message to the receiver in multiple \nbcr{3}{\emph{batches}} of M bits, where the value of each bit is encoded as interference in the row buffer of a DRAM bank. In this attack, we encode logic-1 as interference (i.e., a row buffer conflict), and logic-0 as no interference (i.e., a row buffer hit). 

\figref{fig:pnm_atack} and Listing~\ref{code:pnm-attack} show the end-to-end flow of \attackvone{} and all the steps involved in establishing the covert channel between the sender and the receiver.

\begin{figure}[hb]

    \centering
    \includegraphics[width=\linewidth]{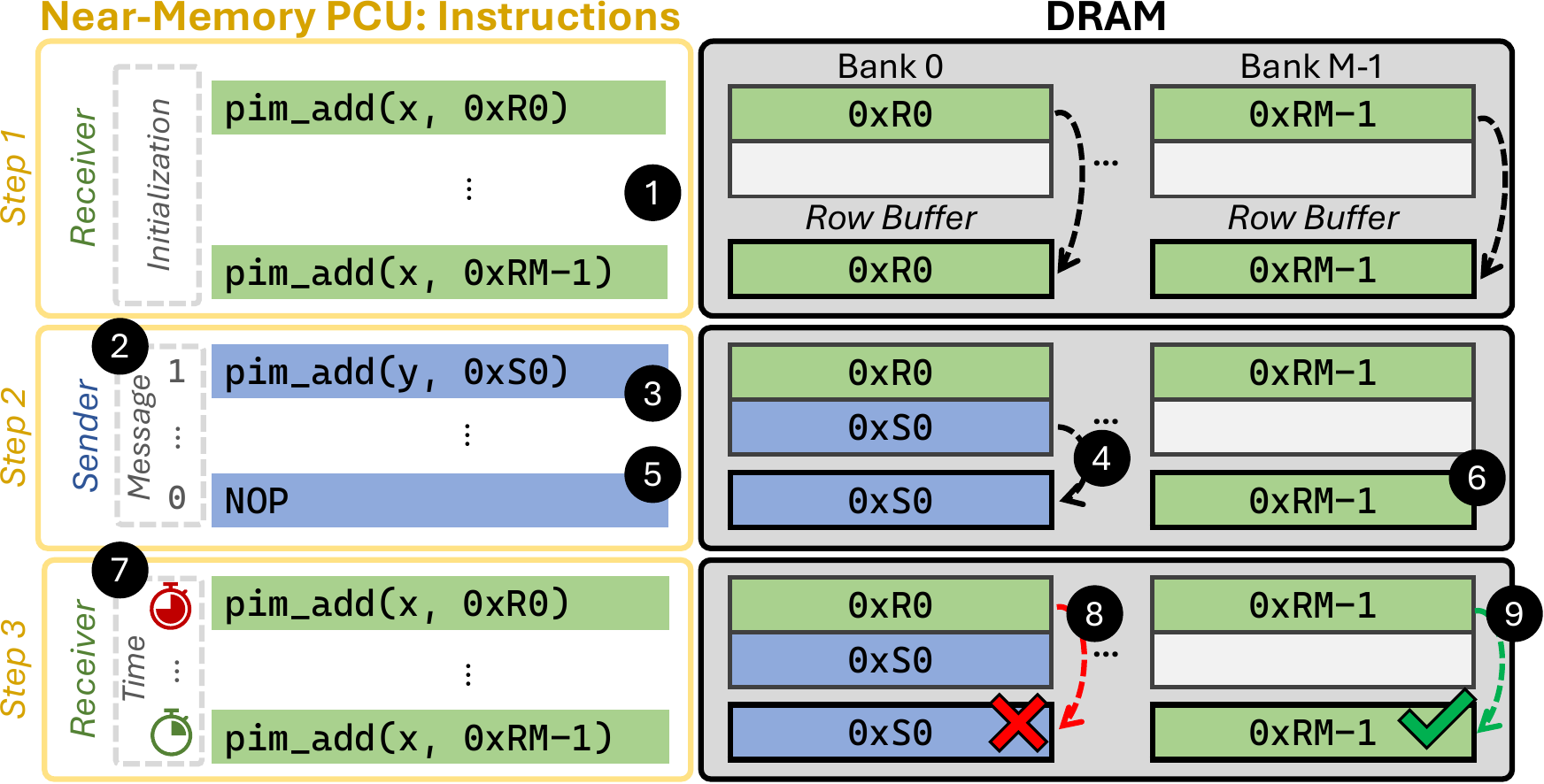}
    \caption{Flow of the PnM-based Covert Channel Attack.}
    \label{fig:pnm_atack}

\end{figure}

\begin{figure}[ht!]
    \begin{lstlisting}[style=customc, label=code:pnm-attack, caption = Pseudocode of PnM-based Covert-Channel Attack., belowskip=-1.55\baselineskip, tabsize=1]
    *@\textbf{Receiver (Step 1)}:@*
    init_row = initialize_DRAM_rows_with_PEIs();

    *@\textbf{Sender (Step 2)}:@*
    message[0: N-1]; // N-bit message
    batch_size = M;  // M-bit batch, where M<N
    for every batch:
        bank = 0;
        for each bit in batch_size;
            if message[bank] == 1: // activate a row
                pim_add(sender_row[bank]);
            else: // do not interfere with the receiver
                NOP();
            bank++;
        // ensure all memory requests are served 
        // before starting the next batch
        memory_fence(); 
        
    *@\textbf{Receiver (Step 3)}:@*
    for every batch:
        bank = 0; 
        for each bit in batch_size;
            timer(start);
            pim_add(init_row[bank]);
            timer(end);
            latency = end - start; 
            if latency > THRESHOLD: // row buffer conflict 
                result = '1';
            else: // row buffer hit
                result = '0';
            bank++;
        memory_fence(); // complete the batch
    \end{lstlisting}
    %\vspace{-5mm}
    \end{figure}

\noindent\textbf{Step 1.} The receiver issues PEIs to initialize each DRAM bank by activating a predetermined row \nbcr{3}{(Line 2, \circled{1}). \nbcr{3}{Initialization copies the receiver's data to row buffers.}}

\noindent\textbf{Step 2.} The sender \nbcr{2}{transmits} a batch \nbcr{3}{of the message} by \nbcr{2}{mapping each bit in the batch to a separate DRAM bank \nbcr{3}{(\circled{2})}. Then, the sender checks the bit value (Line 10).} \nbcr{3}{If the bit is 1, the sender issues a PEI to activate a row (Line 11, \circled{3}) and thus, copies its content to the row buffer (\circled{4}). Otherwise, the sender issues a NOP instruction (Lines 12-13, \circled{5}) and does not copy any new data to the row buffer (\circled{6})}.

\noindent\textbf{Step 3.} The receiver \nbcr{3}{probes} the DRAM banks by issuing PEIs to \nbcr{3}{the initialized rows and measuring the PEI latencies (Lines 23-25, \circled{7}).} 
If the latency \nbcr{3}{of a PEI} exceeds a predetermined \nbcr{3}{row buffer} hit threshold, the receiver \nbcr{2}{detects} interference and decodes the bit as a 1 (Lines 27-28, \circled{8}).
Otherwise, the access is considered a row buffer hit, and the receiver decodes the bit as a 0 (Lines 29-30, \circled{9}).
After the transmission of a batch \nbcr{3}{of the message} (i.e., M bits), the sender and the receiver execute a memory fence \nbcr{3}{(Lines 17 and 32)} to ensure that all the PEIs are executed \nbcr{2}{before moving on to the next batch}. 

\nbcr{2}{The receiver accesses the next cache line in the initialized row to ensure its PEI is executed near memory by bypassing the locality monitor. If the number of batches is greater than the number of cache lines in a row, it uses another row in the same bank to measure access latency.}
The sender and the receiver repeat the same steps for each batch until the entire message is transmitted.

\textbf{Sender-Receiver Synchronization.} 
The sender and the receiver synchronize to 1) ensure their actions do not interfere with each other and 2) overlap the latencies of their operations to increase the throughput of the attack.
In our proof-of-concept implementation, the sender and the receiver use a semaphore. The semaphore's value indicates the number of batches that the sender has completed transmitting, but the receiver has not yet probed. The receiver blocks on the semaphore until the sender increments it (i.e., does not start probing).
When the sender completes the transmission of a batch, they increment the semaphore to signal the receiver to start.
Upon receiving the signal, the receiver decrements the semaphore and starts probing the DRAM banks. 
When the receiver completes probing, they read the semaphore again, and the process repeats until the entire message is transmitted. 
\subsection{PuM-based \X{} Covert-Channel Attack}
\label{sec:pumattack}

\head{Baseline PuM Architecture} 
\revc{\label{resp:c3}3}\revct{We assume a PuM architecture that provides user applications with RowClone~\cite{seshadri2013rowclone} \nbcr{3}{which} enables \nbcr{2}{bulk data copy and initialization operations} within DRAM’s subarray boundaries by exploiting the internal operations of DRAM.
} 
RowClone's Fast Parallel Mode \iey{works by issuing \nbcr{3}{two} activation commands \nbcr{3}{back-to-back}: the first activation copies the content of a source row (i.e., \textit{src}) into the row buffer. The second connects the destination row (i.e., \textit{dst}) to the bitlines. Since the row buffer stores the \textit{src}'s data \nbcr{3}{after the first activation}, by the time the \textit{dst} is activated, the data in \textit{dst} is overwritten by \textit{src}'s data.}

To execute RowClone, we assume that the userspace application specifies \nbcr{2}{1) a source virtual address range, 2) a destination virtual address range, and 3) a mask to copy specific parts of the source \nbcr{3}{range} to the destination \nbcr{3}{range}.} This way, the memory controller receives one RowClone request and breaks it into parallel  RowClone requests, one for each set bit of the mask.

\head{Attack Overview} \nbcr{3}{Our} 
PuM-based covert channel attack (\attackvtwo) exploits in-DRAM bulk copy operations to establish a covert channel between a sender and a receiver. 
\nbcr{2}{Before the attack, the sender and the receiver co-locate their data in the same \nbcr{3}{set of} DRAM banks.
The sender transmits a message to the receiver in multiple \emph{batches} of M bits.
To do so, the sender issues RowClone operations by carefully selecting 1)~source and \nbcr{3}{destination} address ranges corresponding to physical pages spanning \nbcr{3}{the selected set of} DRAM banks, and 2)~a mask to cause \nbcr{3}{interference} only in selected DRAM banks.}
\nbcr{2}{The sender and receiver processes synchronize with barriers similar to \attackvone{} (\secref{sec:pnmattack})}

\figref{fig:pum-attack} and Listing~\ref{code:pum-attack} show the end-to-end flow of \attackvtwo{} and the steps for establishing the covert channel between the sender and the receiver.

\begin{figure}[ht!]
    \centering
    \vspace{1mm}
    \includegraphics[width=0.85\linewidth]{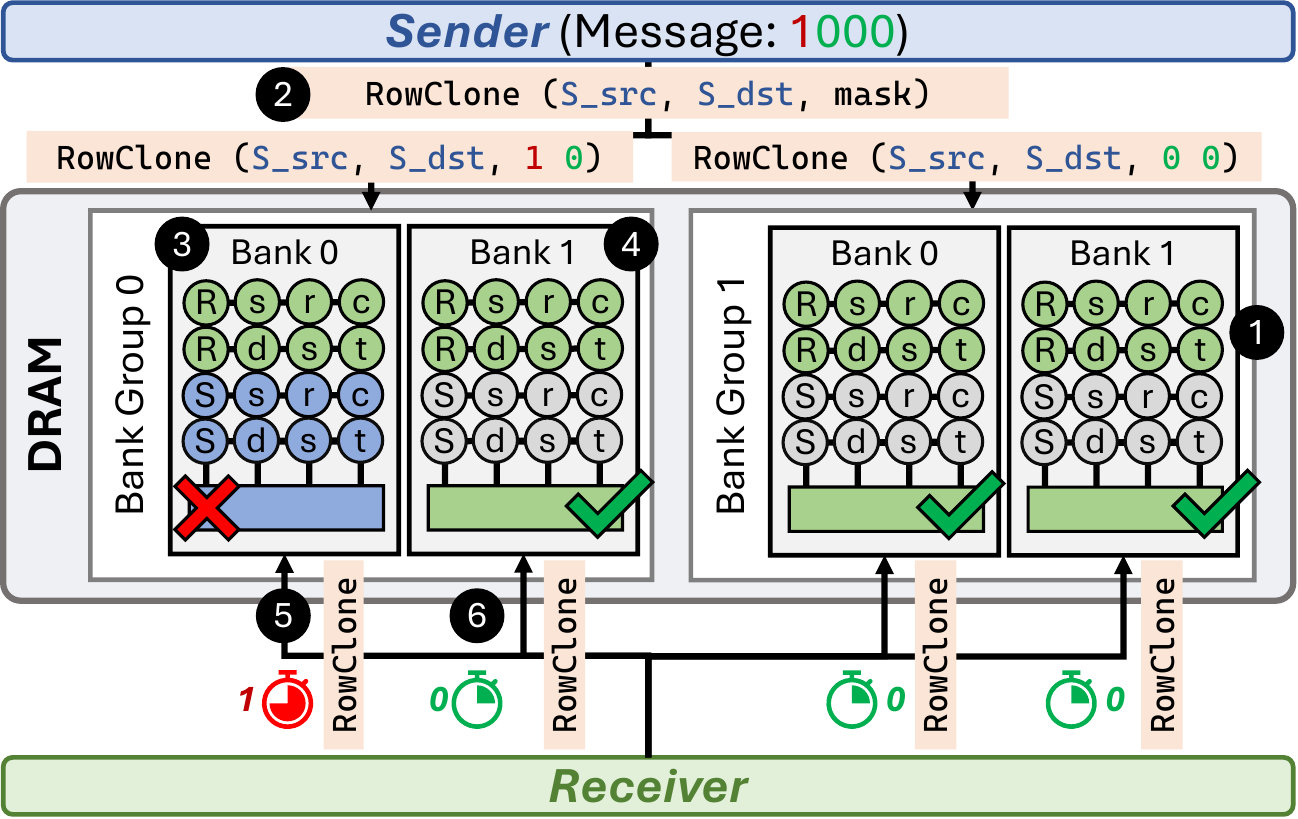}
    \caption{\atb{PuM covert-channel attack flow.}}
    \label{fig:pum-attack}
\end{figure}

\noindent\textbf{Step 1.} First, the receiver initializes \nbcr{3}{the} DRAM banks by issuing a RowClone operation 
\nbcr{3}{(Line 25, \circled{1})}. 

\noindent\textbf{Step 2.} The sender \nbcr{2}{transmits} the message by dividing the message into  M-bit batches. Then, the sender maps each bit of a batch to a separate DRAM bank and creates a mask for the batch \nbcr{3}{(Line 18)}, determining which banks will be executing a RowClone operation (\nbcr{3}{Line 18}).
By using a mask, the sender uses only a single RowClone operation to transmit an M-bit batch \nbcr{3}{of the message} (where M is the number of banks) \nbcr{3}{(Line 20, \circled{2}). The memory controller issues individual RowClone operations to each DRAM bank based on the specified mask (Lines 1-9).} 
To send a 1, the sender triggers RowClone in the corresponding DRAM bank and activates a different row \nbcr{3}{from} the receiver's row in the bank \nbcr{3}{(Lines 4-7, \circled{3})}. 
To send a 0, the sender does not trigger RowClone in the corresponding DRAM bank and does not \nbcr{3}{interfere with the receiver's activated row (Lines 8-9, \circled{4})}. 

\noindent\textbf{Step 3.} The receiver detects interference in DRAM banks using RowClone operations as follows.
The receiver issues one RowClone operation for each DRAM bank \nbcr{3}{to precisely measure the time it takes to complete the operation in each bank \nbcr{3}{(Lines~29-33).} To detect if the sender activated another row, the receiver uses its initial destination address range as the source (i.e., swapping the direction of the copy operation) (Line 31).}
If the sender induces \nbcr{3}{interference} in a DRAM bank, the receiver observes high latency for the RowClone operation, 
\nbcr{3}{indicating} the corresponding bit in the message is set to 1  \nbcr{3}{({Lines~34-35}, \circled{5})}. Otherwise, the receiver detects a fast RowClone operation, which 
indicates that the bit is set to 0 \nbcr{3}{(Lines~36-37, \circled{6})}.

\begin{figure}[!h]
\begin{lstlisting}[style=customc, label = code:pum-attack, caption = Pseudocode of PuM-based Covert-Channel Attack., belowskip=-1.55\baselineskip]
    *@\textbf{Memory Controller}:@*
    row_clone(src, dst, mask): 
        for each DRAM bank:
            if mask[bank] == 1:
                // copy the corresponding src row
                // to the destination row
                row_clone(src, dst, bank);
            else:
                NOP(); // do not interfere with the receiver

    *@\textbf{Sender}:@* 
    message[0:N-1];  // N-bit message
    batch_size = M;  // M-bit batch, where M<N
    start = 0;
    for every batch:
        end = start + batch_size;
        // create mask for the batch
        mask = message[start:end]; 
        // issue a RowClone operation using the mask
        row_clone(src, dst, mask); 
        start = end; // update start for the next batch
        memory_fence(); // complete the batch

    *@\textbf{Receiver}:@*
    init_DRAM_rows_with_RowClone();
    for every batch:
        for each DRAM bank:
            mask = 1 << bank;
            timer(start);
            // issue RowClone to one bank at a time
            row_clone(dst, src, mask); 
            timer(end);
            latency = end - start;
            if latency > THRESHOLD: // row buffer conflict
                result = '1';
            else: // row buffer hit
                result = '0';
        memory_fence(); // complete the batch
\end{lstlisting}
%\vspace{-8mm}
\end{figure}

\head{Advantage over \attackvone{} \nbcr{3}{(\secref{sec:pnmattack}})}
In \attackvtwo{}, the sender can use a single RowClone operation to transmit an M-bit message to the receiver \textit{in parallel}, where M is the number of DRAM banks. Thus, \attackvtwo{} requires
\nbcr{3}{fewer} computational resources to transmit the same amount of information compared to \attackvone{}.
For example, using a single thread, the sender can activate all M DRAM banks and transmit M bits of information in parallel. In contrast, in \attackvone{} \nbcr{3}{(\secref{sec:pnmattack})}, the sender needs to execute one PEI \nbcr{3}{for every} bit.

\subsection{\nba{PnM-based Side-Channel Attack on \nbcr{2}{Genomic} Read Mapping}}
\label{sec:scrm}

\nba{We build a PnM-based side-channel attack that exploits PiM operations to leak private information 
of a concurrently running application. To demonstrate the \X{} side-channel attack, we target a DNA sequence analysis application.}

\head{\nbcr{2}{Genomic} Read Mapping}
{DNA is \nbcr{3}{a} unique identifier of an individual~\cite{arshad2021analysis}, such that even a small 
fraction of DNA can include sufficient information for linking \nbcr{2}{a DNA sample} to the owner's identity~\cite{lin2004genomic, lu2021methods}. 
Therefore, preserving privacy \zb{while processing genomic data} carries utmost importance \zb{to protect sensitive information}. 
\zb{Sequencing the} DNA of a biological sample produces short DNA fragments called \textit{reads}~\cite{shendure2017dna}. \zb{The reads are then processed by} 
\textit{read mapping}~\cite{alser2020accelerating,alser2021technology,alser2022from,lin2004genomic,lu2021methods,wang2009learning,arshad2021analysis}, \zb{which} is a fundamental \nbcr{3}{task performed} \nba{to find} the matching locations of 
reads in the reference genome that acts as a template to construct the complete genome of the sample.}

\nba{Read mapping (RM) has two key steps: (i) seeding~\cite{berlin2015assembling} and (ii) alignment~\cite{delcher1999alignment,Li2013AligningSR}.  \figref{fig:dna-read} shows a high-level overview of RM. 
During the seeding step, RM identifies the possible locations of the reads \nbcr{3}{in} the reference genome based on sequence similarities 
between the reads and the reference genome. Seeding involves hashing small segments of the DNA sequences 
(i.e., seeds) \nbcr{3}{(\circled{1})} and probing a hash table that is built from the reference genome \nbcr{3}{(\circled{2})}. \nbcr{3}{The hash table provides potential locations for the seed in the reference genome (\circled{3}).} \nbcr{3}{In the alignment step}, RM aligns the reads with 
their candidate regions \nbcr{3}{in} the reference~(i.e., potential reference segments) using dynamic programming methods~\cite{alser2021technology,lin2004genomic} to find each read's best matching 
\zb{position} among all possible locations \nbcr{3}{(\circled{4})}.}

\begin{figure}[h!]
    \centering
    \includegraphics[width=0.9\linewidth]{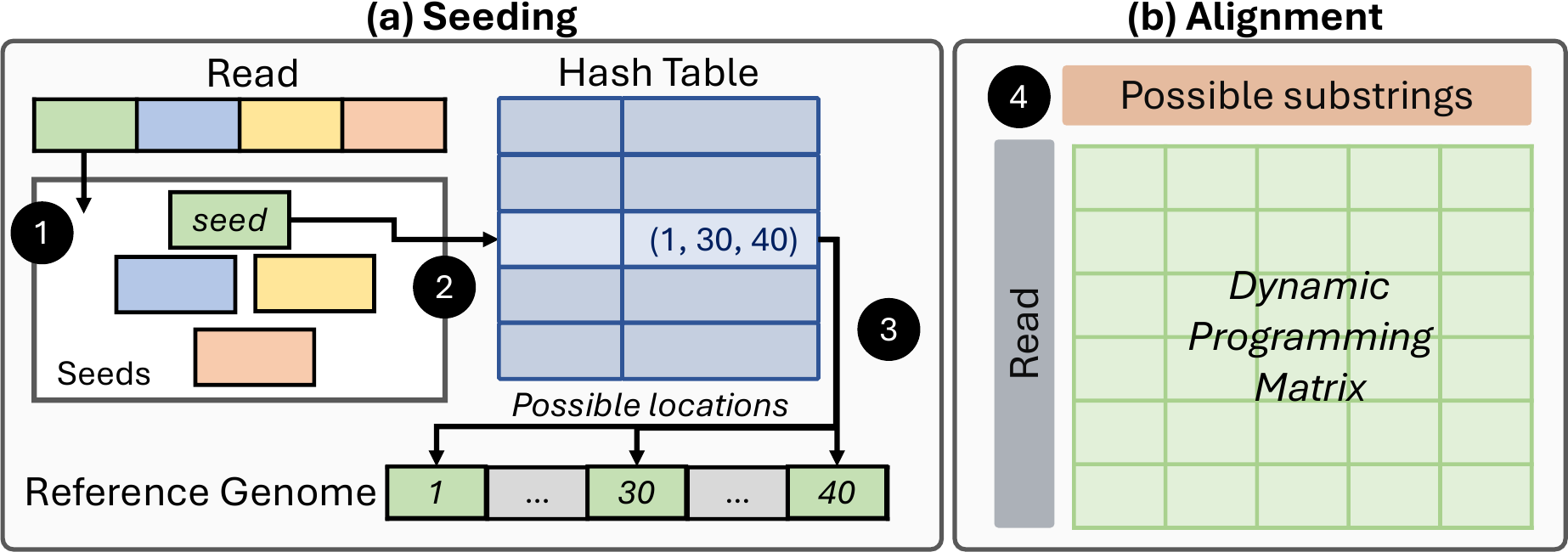}
    \caption{DNA Read Mapping Operation: Seeding and Sequence Alignment.}
    \label{fig:dna-read}
\end{figure}

\cqt{RM requires a large amount of data movement to access different parts of the hash table and corresponding regions on the reference genome. 
Hence, several prior works aim to improve the performance of RM by offloading seeding and alignment to PiM-enabled systems 
to reduce the data movement overheads~\cite{kim2018grim,huangfu2018radar,khatamifard2021genvom, cali2020genasm, gupta2019rapid,li2021pim,angizi2019aligns,zokaee2018aligner,singh2021fpga,diab2023framework,alonso2024bimsa}.
In this work, we demonstrate that \nba{an attacker can launch a high-throughput side-channel attack against RM {by utilizing PiM instructions}} 
and quickly leak a large amount of information about the \nbcr{3}{query} genome.}

\head{Attack Overview}
We assume that the victim is running \nbcr{3}{an} RM application on a PiM-enabled system and the attacker is able to run a malicious application on the same system. 
The attacker and the victim are using the same read mapping tool (e.g., minimap2~\cite{li2018minimap2}).
\nba{The read mapping tool constructs a hash table that contains information about the seed locations in the reference genome, 
and allows each user in the system to probe the hash table via queries during read mapping. We assume the hash table is distributed 
across multiple DRAM banks. This assumption is realistic, as many modern DRAM address mapping schemes 
{(e.g.,~\cite{hur2019adaptive,kaseridis2011minimalistic,liu2018get,ghasempour2016dream})} 
aim to interleave consequent memory chunks across different DRAM banks to exploit bank-level parallelism~\cite{mutlu2008parbs,salp}.} 

\figref{fig:dnattack} shows the high-level overview of the attack. \nbcr{3}{The hash table is distributed across banks as explained above (\circled{1}). During the attack,} the victim application \zb{extracts seeds from the reads}
and offloads the corresponding seeding and alignment \nbcr{3}{steps} to the PiM-enabled system (\circled{2}). 
During the seeding step, the corresponding DRAM row gets activated (i.e., Index 0-M of the hash table) (\circled{3}).
The attacker continuously uses PnM instructions to probe the shared hash table in an attempt to identify if the victim application is accessing 
a specific index of the hash table (\circled{4}). 
If the attacker observes that the victim application is accessing a specific index, it retrieves partial or exact information about the 
potential locations of the \zb{reads} in the reference genome.
As shown in previous works on DNA imputation~\cite{impute1,impute2, deznabi2017inference, humbert2017quantifying}, 
the attacker can use the leaked information \zb{in a completion attack~\cite{lu2021methods}} to infer properties about 
some regions of the private \nbcr{3}{query} genome (\circled{5}).

\begin{figure}[h!]
    \centering
    \includegraphics[width=0.9\linewidth]{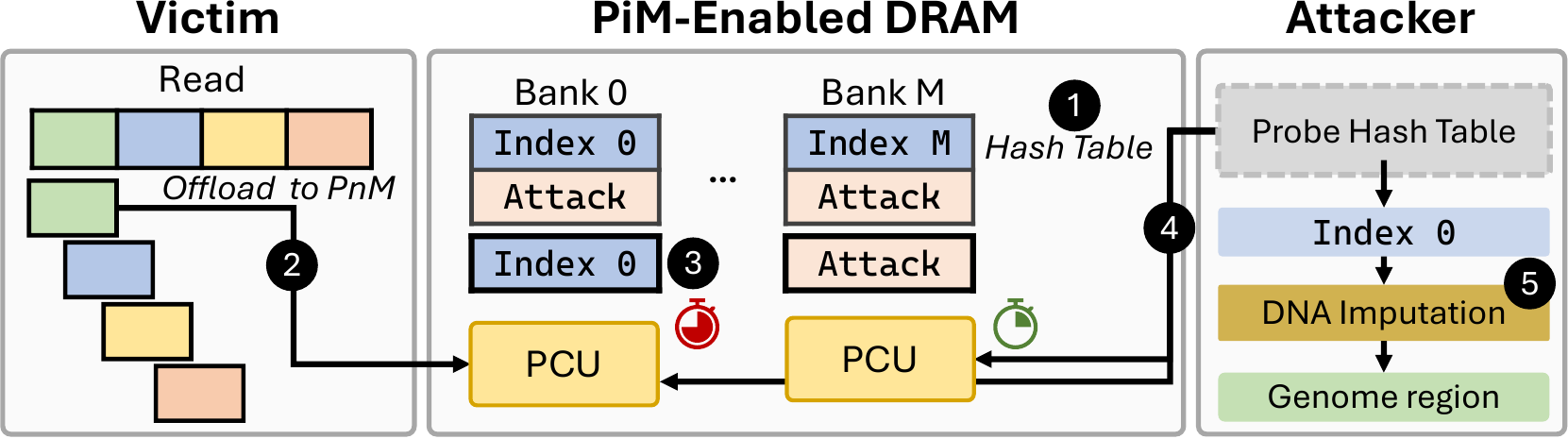}
    \caption{\nba{High-level overview of \nbcr{3}{the} \X{} attack model on PiM-enabled \nbcr{3}{genomic read mapping}.}}
    \label{fig:dnattack}
    % \vspace{-4mm}
\end{figure}

\setcounter{version}{4}

\section{Methodology}

\subsection{Threat Models}
\label{sec:method}

\head{\nba{Covert Channel} Threat Model}
We assume a scenario where a sender and a receiver operate on the same PiM-enabled system to exchange information. 
PiM instructions are available to userspace applications without any restrictions. 
The receiver has access to \texttt{cpuid} and \texttt{rdtscp} instructions~\cite{intelmanual2016}, enabling high-precision measurement of memory access latencies.
For \attackvone{}, the sender and the receiver co-locate data in the same \nbcr{3}{set of} DRAM banks.
\nb{For \attackvtwo{}, the sender allocates two virtual address ranges whose corresponding physical pages span across \nbcr{3}{the same set of} DRAM banks.} 

\head{\nba{Side Channel Threat Model}} \nba{We assume a scenario where an attacker process 
operates on the same PiM-enabled system as a victim process and leaks information about the 
victim process. We use a \nbcr{2}{genomic read mapping} implementation based on minimap2~\cite{li2018minimap2}, 
and we modify the implementation to offload the seeding and alignment steps to \nbcr{3}{the} PiM-enabled 
system~\cite{PEI}. We assume that the alignment step includes chaining~\cite{alser2021technology,li2018minimap2}.  We use the human genome as the 
reference and compare it against synthetic \nbcr{3}{query} genomes. We experiment with multiple seed 
sizes and report the best-performing selection.}

\subsection{Evaluation Methodology}

\subsubsection{\rvb{Proof-of-Concept}}
We construct a proof-of-concept of \nba{all} \X{} variants using the Sniper Multicore Simulator~\cite{sniper}.\footnote{There is no commercially available system that follows a similar architecture as PEI or any modern system that enables end-to-end RowClone operations.} 
We extend Sniper to (i) accurately model all internal DRAM operations, 
(ii) emulate and simulate the PiM interface and PiM operations, and 
(iii) emulate the functionality of the \texttt{cpuid} and the \texttt{rdtscp} 
instructions~\cite{intelmanual2016} to measure execution \nbcr{3}{time}. 

Table~\ref{tab:simconfig} shows the configuration of our simulation setup.
Before launching the attack, the sender and the receiver threads warm up to avoid compulsory DRAM accesses for instructions or page table walks. 
\nbcr{3}{For the \gls{pnm} system~\cite{PEI}, we model the additional latency of a PEI (e.g., latency of accessing additional system structures) as $3$ cycles~\cite{PEI}.}

\begin{table}[h]
    \centering
        \caption{\nbcr{3}{Simulated System} Configuration
        }
    \scriptsize
    \begin{tabular}{m{6.5em} m{26em}}
    \toprule
    \textbf{CPU} & 4-core, 4-way \nbcr{3}{issue}, OoO x86 core, 2.6GHz\\
    \midrule
    \textbf{MMU} &L1 DTLB (4KB):  64-entry, 4-way, 1-cycle, L1 DTLB (2MB): 32-entry, 4-way, 1-cycle, L2 TLB: 1536-entry, 12-way, 12-cycle \\        
    \midrule
    \textbf{L1 Cache} & I-cache: 32~KB; L1D Cache: 32~KB, 8-way, 4-cycle, LRU, \revbt{IP-stride prefetcher~\cite{stride}} \revb{1}\\
    \midrule
    \textbf{L2 Cache} & 2~MB, 16-way, 16-cycle, SRRIP~\cite{srrip}, \revbt{Streamer \nbcr{3}{prefetcher}~\cite{streamer}}\\
    \midrule
    \textbf{L3 Cache} & 2~MB/core, 16-way, 50-cycle, SRRIP~\cite{srrip} \\ 
    \midrule
    \textbf{Main Memory} & DDR4-2400, 16 Banks, 4 bank groups, 1 Rank, 1 Channel, Row size = 8192 bytes, $t_{RCD}$ = 13.5 ns, $t_{RP}$ = 13.5 ns , $t_{RC}$ = 13.5 ns, Open Row policy, Row Timeout = 100 ns\\ 
    \bottomrule
    \end{tabular}

    \label{tab:simconfig}
\end{table}

\subsubsection{\nba{\rvb{Comparison Points for Covert-Channel Attacks}}} To showcase the effectiveness of \X{}, we evaluate the leakage throughput of five different covert-channel attacks.

\noindent(i) \textbf{DRAMA-clflush:} The row buffer-based covert-channel attack proposed in~\cite{pessl2016drama} which uses \texttt{clflush} instructions~\cite{intelmanual2016} to send memory requests directly to main memory. \revd{\label{resp:d4}3}{\texttt{clflush} only probes the LLC to flush the cache line.} 

\noindent(ii) \textbf{DRAMA-Eviction:} The row buffer-based covert channel attack proposed in~\cite{pessl2016drama}, which uses cache eviction sets~\cite{liu2015last} to send memory requests directly to main memory.

\noindent{(iii) \textbf{DMA Engine:} A row buffer-based attack that uses the DMA engine~\cite{arm2000dma,microchip2008direct} to bypass the cache hierarchy.

\noindent(iv) \textbf{\attackvone{} (\secref{sec:pnmattack}):} \atb{An} attack that bypasses the locality monitor and offloads PEI instructions directly to compute units near DRAM banks in order to establish a covert channel. 

\noindent(v) \textbf{\attackvtwo{} (\secref{sec:pumattack}):} \atb{An} attack that leverages RowClone to offload bulk-copy operations directly to main memory and establish a covert channel. 

\subsubsection{{\rvb{Noise Sources in Simulation Setup and Error Rate}}}
\nba{We simulate hardware prefetchers~\cite{streamer,stride} and page table walkers~\cite{intelmanual2016} to induce noise in the simulated system.}
\revbt{We measure the throughput of each attack only based on the successfully leaked data (i.e., incorporating error rate in our measurements).}

\section{Evaluation}
\subsection{Validation of Proof-of-Concept}
\nba{We \nbcr{4}{first demonstrate proof-of-concept attacks for \attackvone{} and \attackvtwo{}.}} 
\nbcr{2}{We send 16-bit messages with a batch size of 16 (i.e., using 16 banks in parallel).}
\zb{\figref{fig:poc_pnm}} shows the latency (in cycles) measured by the receiver \nbcr{4}{during transmission of example messages. The y-axis shows the measured latency} for each (a) PEI and (b) RowClone operation for \attackvone{} and \attackvtwo{}, respectively. \nbcr{2}{The x-axis shows the transmitted bit values \nbcr{3}{where each bit is transmitted using a different bank.}} 
\nbcr{2}{We make two key observations. First, for both attacks,}
\nb{the receiver successfully \nbcr{3}{determines} the \nbcr{4}{transmitted} bit values by detecting \nbcr{2}{the row buffer conflicts caused by the sender}.}
\nbcr{2}{Second, }
the receiver \nbcr{3}{decodes} the complete message using a threshold value of 150 cycles.
If the \nbcr{2}{measured} latency is less than 150 cycles (i.e., row buffer hit), the receiver decodes a bit as logic-0; otherwise (i.e., row buffer conflict), it decodes the bit as logic-1.

\begin{figure}[h!]
    \centering
    \includegraphics[width=0.9\linewidth]{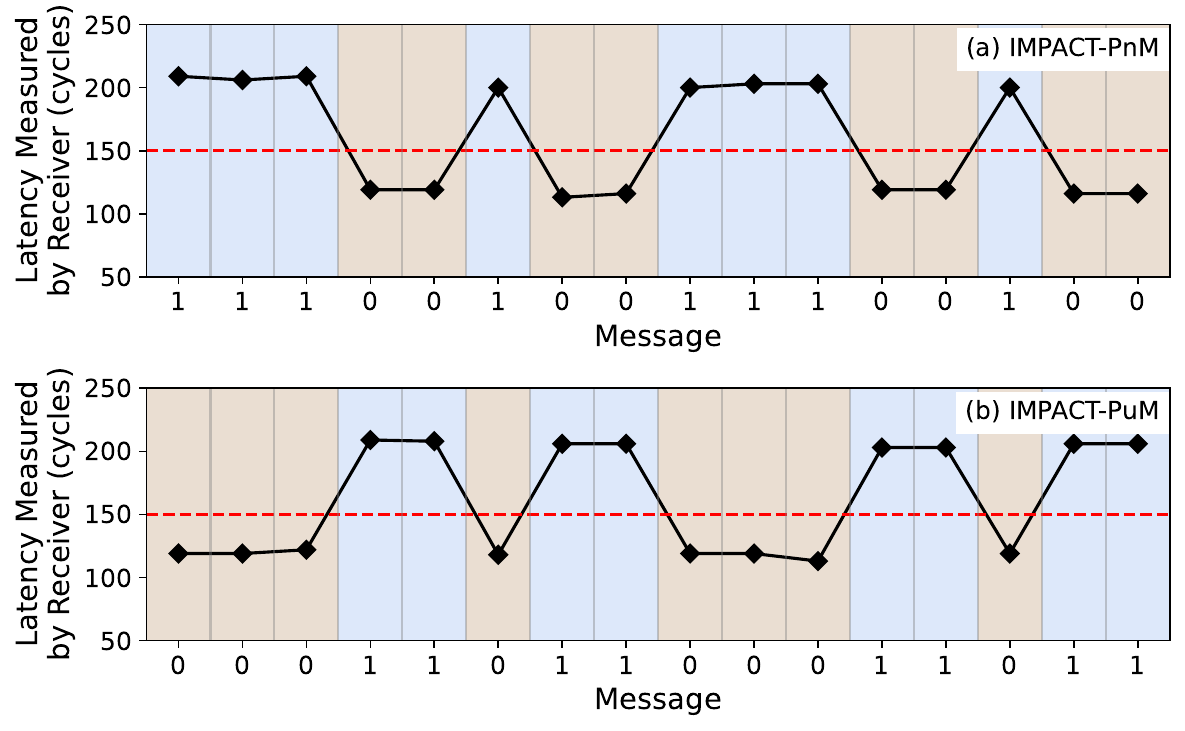}
    \caption{Latency (in cycles) measured by the receiver to execute (a) a PEI and (b) a RowClone \nbcr{3}{operation} for each bank to decode a 16-bit message.}
    \label{fig:poc_pnm}
\end{figure}

\subsection{\nbcr{3}{Covert Channel} Throughput Analysis}
\label{sec:analysis}

\figref{fig:throughput} shows the throughput achieved by \attackvone{}, \attackvtwo{} and \nbcr{3}{state-of-the-art main memory-based} covert-channel attacks \nbcr{3}{in modern systems with increasing LLC sizes}. \nbcr{3}{The x-axis shows the increasing LLC size (from 1 MB to 128 MB), and the y-axis shows the leakage throughput of the covert channels.}  \nbcr{3}{For all attacks,} 
\nb{one sender and one receiver process run concurrently} \nbcr{3}{and transmit a fixed-length message.}

We make \param{four} key observations. First, \attackvone{} and \attackvtwo{} achieve significantly higher throughput 
than all other attack vectors. \nba{\attackvone{} and \attackvtwo{} \aedraft{\nbcr{4}{provide} communication throughputs of 
\param{8.2} Mb/s and \param{14.8} Mb/s, respectively, irrespective of the cache size. They support up to 
\param{3.6$\times$} and \param{6.5$\times$}} of the throughput achieved by the state-of-the-art main memory-based attack, DRAMA-clflush.} 
Second, \attackvtwo{} \aedraft{\nbcr{4}{provides} 82\% higher throughput than \attackvone{}. \nbcr{3}{This is because the sender in \attackvtwo{} can transmit a message with lower latency using fewer operations (\secref{sec:pumattack}).}} 
Third, as the LLC size increases, the throughputs of DRAMA-eviction and DRAMA-clflush decrease significantly \nbcr{3}{due to the increasing LLC access latency}.
\revat{Fourth, the DMA engine attack \nbcr{4}{provides} \aedraft{a throughput of 0.81 Mb/s irrespective of the LLC size, \nbcr{3}{which is} \param{10.1}$\times$ slower than \attackvone{} due to its OS-related overheads} (e.g., context switch overhead, OS instructions).} 
We conclude that \attackvone{} and \attackvtwo{} achieve significantly higher throughput than the existing state-of-the-art main memory-based covert-channel attacks.

\begin{figure}[ht!]
    \centering
    \includegraphics[width=\linewidth]{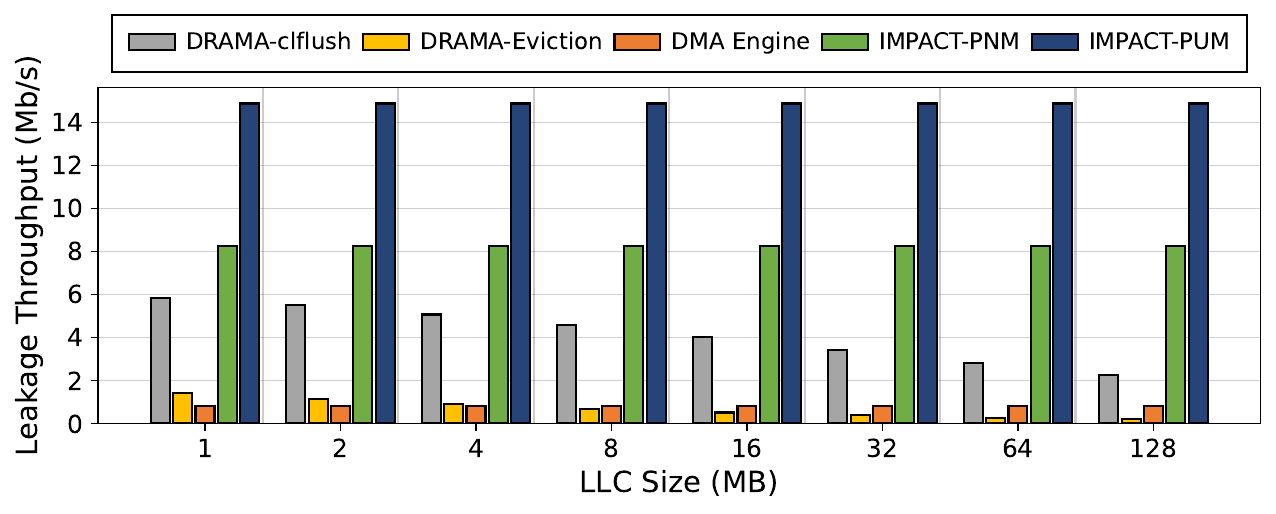}
    \caption{\nbcr{4}{Leakage throughput of \attackvone{}, \attackvtwo{}, and state-of-the-art main memory-based covert channel attacks.}}
    \label{fig:throughput}
\end{figure}

\head{\revdt{Breakdown of Throughput Improvements}}\revd{\label{resp:d3}2}
\label{sec:breakdown}
\nbcr{2}{We analyze the throughput breakdown of \attackvone{} and \attackvtwo{} by calculating the time spent on sender and receiver routines to transmit a fixed-length message.} 
\figref{fig:send_rev} shows the time (in CPU cycles) it takes for the sender to send the message and for the receiver to 
decode it in IMPACT-PnM and IMPACT-PuM. We \nbcr{2}{make two key observations. First,} in \attackvtwo{}, the sender routine takes 11.1$\times$ less time than the sender in \nbcr{4}{\attackvone{}}. This is because \attackvtwo{} can issue \nbcr{4}{\emph{only}} one RowClone request \nbcr{2}{to transmit N bits, where N is the number of banks. In contrast, \attackvone{} transmits the same message with potentially multiple PEIs.
Second, the receiver routines \nbcr{4}{of \attackvone{} and \attackvtwo{}} spend similar amounts of time decoding the message as they both need to use distinct \gls{pim} operations to detect which DRAM banks were accessed by the sender.
Based on these observations, we conclude that \attackvtwo{} has a higher throughput than \attackvone{} due to the differences in the sender routines.
}

\begin{figure}[ht!]
    \centering
    \includegraphics[width=0.9\linewidth]{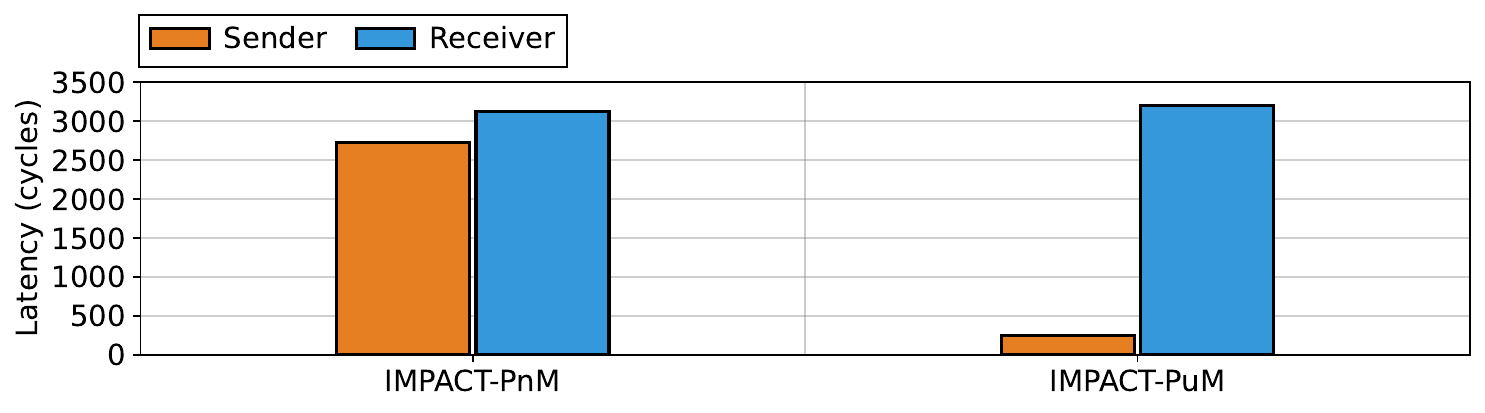}
    \caption{Time (in cycles) it takes (i) for the sender to send the message and (ii) for the receiver to read the message in IMPACT-PnM and IMPACT-PuM.}
    \label{fig:send_rev}
    \vspace{-3mm}
\end{figure}

\subsection{Side-Channel Attack Throughput Analysis}
\label{sec:side-channel-eval}
\nbcr{3}{We evaluate the side-channel attack against genomic read mapping (RM) \nbcr{4}{(\secref{sec:scrm})} in a PiM-enabled system~\cite{PEI}, \nbcr{4}{sweeping} the number of DRAM banks that store the hash table used in the seeding step.}
\figref{fig:dnattack_res} shows the throughput and the error rate of the side-channel attack in the \nbcr{2}{primary y-axis and secondary y-axis}, respectively. \nbcr{3}{The x-axis shows the \nbcr{4}{total} number of DRAM banks in the system.} 
We measure the throughput of the attack based on the correct guesses \nbcr{4}{of the hash table entries accessed} and calculate the error rate based on the number of incorrect guesses.\footnote{The exact end-to-end accuracy of identifying the properties of the \nbcr{3}{query} genome depends on the chaining mechanism and the imputation algorithm~\cite{impute1,impute2} used by the attacker, and it is out of the scope of this work.}

\begin{figure}[h!]
    \centering
    \includegraphics[width=0.85\linewidth]{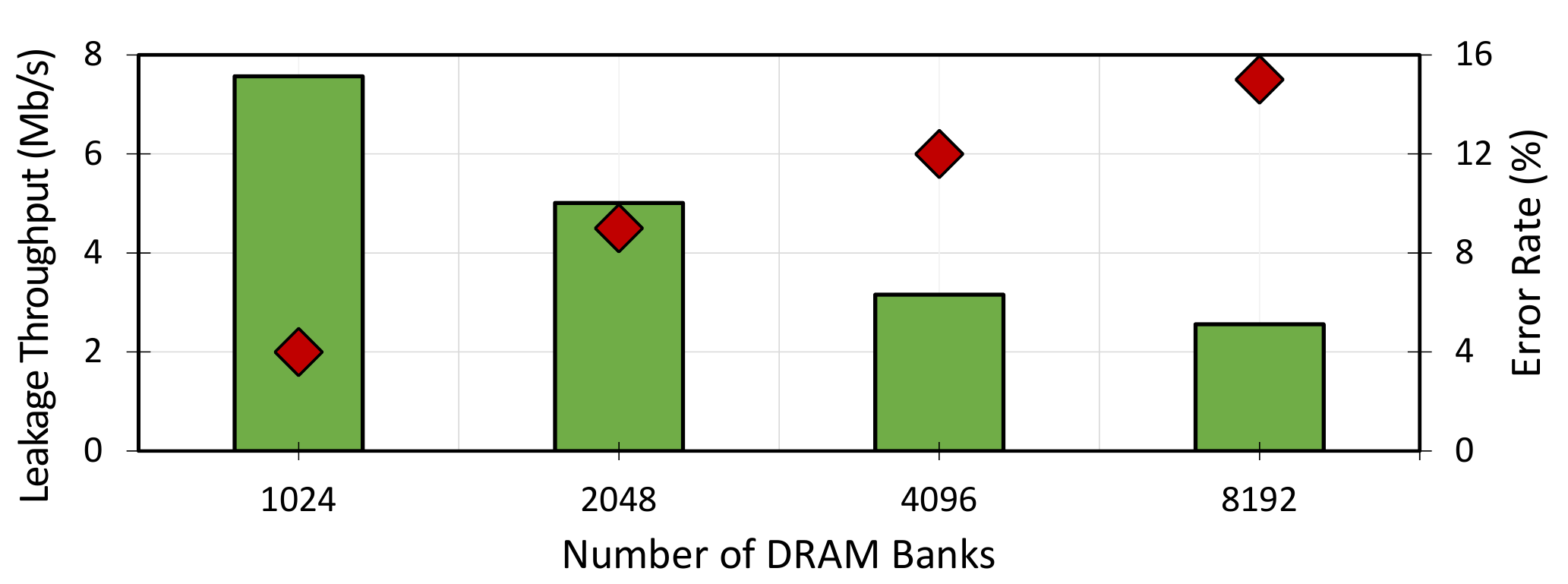}
    \caption{\cqt{Leakage throughput and error rate of the genomic read mapping side-channel attack.}}
    \label{fig:dnattack_res}
\end{figure}

We make \param{two} key observations.
First, the attacker can leak information about the \nbcr{3}{query} genome at a throughput of 7.57 Mb/s and an error rate of $<5$\% \nbcr{3}{in a system with 1024 banks}.
Second, as the number of banks increases, the throughput of the attack decreases to 2.56 Mb/s and the error rate increases to $<15$\%. This is because the attacker has to probe more banks, making the attack more prone to noise and reducing its bandwidth.

However, at the same time, the leaked information provides more exact information about the \nbcr{3}{query} genome since the number of hash table entries per bank decreases, reducing the number of potential locations of the reads in the reference genome per bank.
For example, in a PiM-enabled DRAM device with 1024 banks, the victim activates a row with 16 hash table entries. Assuming the attacker identifies the activated bank correctly, they determine that one of the 16 hash table entries was targeted by the victim. In contrast, in a PiM-enabled device with 2048 banks, the victim activates a row with 8 hash table entries. When the attacker identifies the activated bank correctly, they determine that one of the 8 hash table entries was targeted by the victim, which leaks more precise information on the \nbcr{3}{query} genome.

\nbcr{2}{Based on these observations, we conclude that an attacker can successfully leak \nbcr{3}{query} genome characteristics in a \gls{pim}-enabled system \nbcr{4}{by} leveraging \gls{pim} operations.}

\section{Defenses \nbcr{4}{Against} \X{}}
\label{sec:mitigations}
\rvd{We \nbcr{4}{demonstrate} high-throughput attack vectors that can emerge with the adoption of PiM architectures. 
Our results provide insights into new side and covert channel attack vectors that PiM systems might face, and show that these attack vectors should be considered while designing such PiM systems. In this section, we discuss }
the effectiveness of four defense mechanisms that \nbcr{2}{mitigate} the row buffer-based timing channel \nbcr{4}{we introduce}.

\subsection{\rvb{Memory Partitioning (MPR)}}
\label{sec:mpr}
Bank-level memory partitioning \nbcr{4}{(similarly to~\cite{muralidhara2011,liu2014bpm})}
\nbcr{2}{allocates each DRAM bank to \nbcr{4}{\emph{only}} one process and enforces that only that process accesses the DRAM bank. This way, none of the processes can observe latency differences due to other processes' memory accesses. Thus, the attacker processes cannot build communication channels or leak information.}
\revet{However, this approach has three major drawbacks. First, it limits the number of applications that can run concurrently due to the fixed number of DRAM banks available in the system. Second, it can cause memory underutilization due to allocating DRAM bank-sized memory to applications. Third, it disables data sharing \nbcr{4}{across processes}, which can significantly increase data duplication and memory underutilization, and lead to frequent swapping from the disk when the memory space is full.}

\subsection{\rvb{Closed Row Policy (CRP)}}
\label{sec:crp}
Enforcing a memory controller policy that closes \agy{a} DRAM row after each operation, a \textit{closed row policy} \nbcr{4}{(similar to as described in~\cite{kaseridis2011minimalistic,blackmore2013quantitative,kahn2004method})}, can be an effective defense against \X{}. 
By closing the DRAM row after each operation, the memory controller ensures that subsequent memory accesses cannot access the same row without requiring a row activation.
The closed row policy disrupts the timing patterns exploited by \X{}.
However, it also leads to significant performance overheads \nbcr{4}{\cite{luo2023rowpress,blackmore2013quantitative}}, even for workloads with low data locality, as every DRAM access leads to a row buffer miss.

\subsection{{Constant-Time DRAM Accesses (CTD)}}
\label{sec:ctd}
Another defense mechanism is to enforce a constant-time DRAM access (CTD) policy. \agy{When} a memory access is issued, 
the memory controller waits for a fixed amount of time before returning the data to the processor. 
This fixed amount of time is equal to the worst-case DRAM access latency. 
By enforcing CTD, the memory controller ensures that all memory accesses take the same amount of time, \nbcr{4}{which eliminates} the timing \nbcr{4}{differences} exploited by \X{}. 
However, this leads to significant performance overheads (\secref{sec:mitig_overhead}) as every DRAM access has the worst-case latency.

\subsection{\revcommon{Reducing the Attack Throughput with Adaptive Constant-Time DRAM} \nbcr{4}{(ACT)}}
\label{sec:selective}
\label{sec:mitig_overhead}

\revcommon{Another approach is to reduce the leakage throughput (without completely getting rid of the covert-channel attack)  while incuring lower performance overheads than existing defenses. To understand the effectiveness of such a solution,
we implement a memory controller-based countermeasure that adaptively enforces constant-time memory access latency \nbcr{4}{per DRAM bank} based on the row buffer contention \nbcr{4}{observed}. We call this countermeasure Adaptive Constant-Time (ACT). ACT counts the row buffer conflicts \nbcr{4}{for each DRAM bank} during \nbcr{4}{an} epoch and decides which latency policy (i.e., constant-time or default latency) to employ in the next one. It switches to the constant-time policy \nbcr{4}{for a bank} when there are more row buffer conflicts than a predetermined threshold in the last epoch. This way, in the next epoch, all memory requests being served by the bank will exhibit the same latency (i.e., the worst case latency), and the attacker always observes interference. While employing constant-time latency, ACT still keeps track of the row buffer conflicts. It switches back to the default policy only when the number of conflicts is lower than the threshold in the last epoch. This can reduce the performance overheads caused by \nbcr{4}{CTD}. ACT does not fundamentally solve the row buffer-based timing channel, and an attacker can still mount the attack by transmitting messages \nbcr{4}{only when ACT is employing default latency} and then waiting \nbcr{4}{while ACT employs} constant-time \nbcr{4}{latency}. However, this limits the usable epochs during execution and reduces the attack throughput.}

\revcommon{We implement ACT and evaluate \nbcr{4}{its impact on system performance and the attack throughput}.} We evaluate three different configurations \nbcr{4}{an epoch is set to 1000ns}: (i)~\nbcr{4}{\emph{ACT-Aggressive}}: \nbcr{4}{employing} constant-time \nbcr{4}{latency} for the next 4000 epochs after the 1st conflict occurs in the bank, (ii)~\nbcr{4}{\emph{ACT-Mild}}: \nbcr{4}{employing} constant-time \nbcr{4}{latency} for the next 2 epochs after the 1st conflict occurs in the bank, (iii)~\nbcr{4}{\emph{ACT-Conservative}}: \nbcr{4}{employing} constant-time \nbcr{4}{latency} for the next 2 epochs after 5 conflicts occur in the bank. \nb{\nbcr{4}{To analyze ACT's impact on system performance,} we evaluate four workloads from \nbcr{4}{the} GraphBIG benchmark suite~\cite{Lifeng2015}: Betweenness Centrality, Breadth-First search (BFS), Connected components (CC), Triangle counting (TC) and, XSBench (XS)~\cite{Tramm2014}.} \nbcr{4}{To analyze ACT's impact on attack throughput, we calculate the reduction in \attackvone{}'s throughput.} 
Figure~\ref{fig:mitig_overhead} shows the \nbcr{4}{performance comparison of \textit{ACT-Conservative}, \textit{ACT-Mild}, \textit{ACT-Aggressive}, and} CTD \nbcr{4}{normalized} to \nbcr{4}{a system with no \X{} defense}.

\begin{figure}[h]
    \centering
    \includegraphics[width=0.9\linewidth]{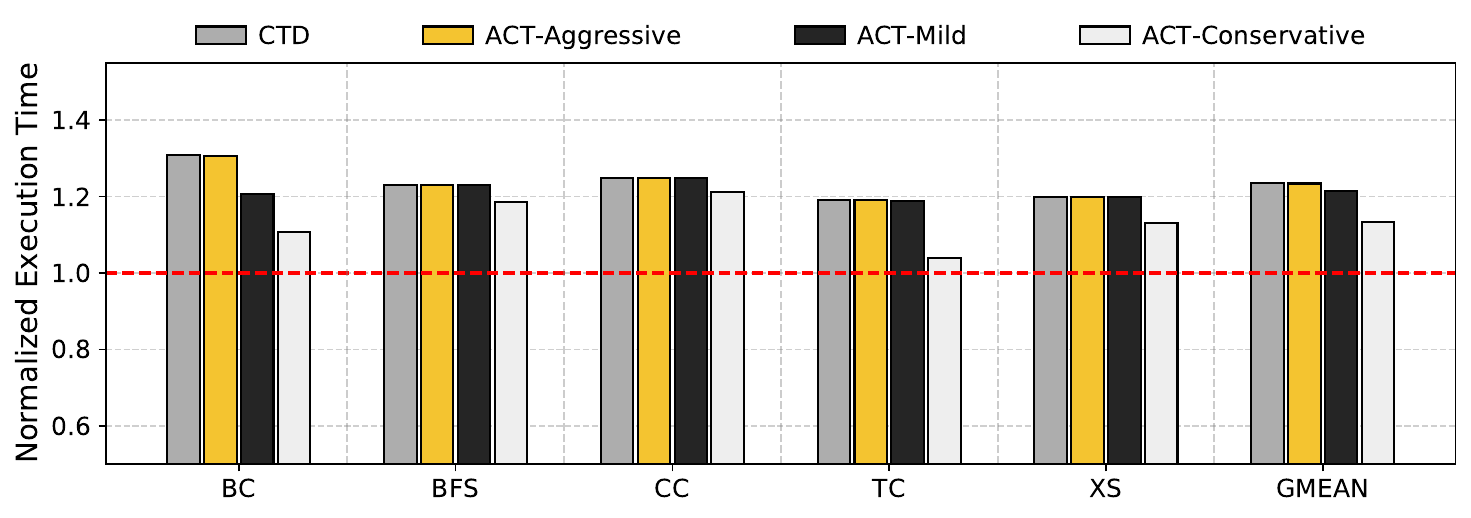}
    \caption{Performance \nbcr{4}{comparison of} Constant-Time policy and Adaptive Constant-Time defense policies over the baseline system \nbcr{4}{with no \X{} defense.}}
    \label{fig:mitig_overhead}
    \vspace{-3mm}
\end{figure}

We make two key observations. First, \textit{ACT-Aggressive} causes similar performance overhead as CTD, while reducing the throughput of \nbcr{4}{\attackvone{}} by 72\% \nbcr{4}{on average (not shown in the figure)}, in contrast to CTD, which prevents the attack completely. Second, the two less aggressive variants of ACT (\textit{ACT-Mild} and \textit{ACT-Conservative}) lead to only 10\% performance overhead. However, they cannot reduce the throughput of \attackvone{}. This is because the attack utilizes all available banks during a batch of transmission, which increases the idle time between each transmission that uses a specific DRAM bank (i.e., the attack uses the same bank only after a time period where it issues requests to the other banks). This requires significantly high delay times for the mitigation to reduce the throughput. Our findings highlight the necessity for a new approach in defending against main-memory timing attacks.

\section{Discussion}

\subsection{\rva{Other Potential PiM-Based Attack Vectors}}\labela{1}
{In this work, we show that emerging PiM architectures can enable a set of high-throughput main memory-based timing attacks based on the DRAM row buffer. Adopting PiM architectures can lead to other attack vectors \nbcr{2}{or exacerbate existing ones,} \nbcr{4}{requiring} further exploration and analysis. \nbcr{2}{One such example is the impact of \gls{pim} on read disturbance vulnerability~\cite{kim2014flipping,luo2023rowpress}. A} prior PiM proposal \nbcr{4}{in 2021}~\cite{hajinazar2021simdram} states that the proposal may lead to increased vulnerability to RowHammer~\cite{kim2014flipping} attacks. \nbcr{2}{\nbcr{4}{A new work in 2025,} PuDHammer~\cite{yuksel2025pudhammer}, studies the read disturbance effects of \gls{pud} operations and shows that these operations \nbcr{4}{greatly} increase read disturbance vulnerabilities of DRAM modules (up to 158.6$\times$).} Thus, it is critical to employ robust and scalable \nbcr{4}{RowHammer~\cite{kim2014flipping} and RowPress~\cite{luo2023rowpress}} solutions \nbcr{2}{to prevent read disturbance bitflips in \gls{pim}-enabled systems.} 

Other potential attack vectors include 1) thermal attacks that leverage PiM operations to increase the temperature in DRAM to make DRAM unavailable and restrict its service time, and degrade reliability, and 2) timing attacks that leverage coarse-grained function offloading. Exploring these directions requires special attention and research, which we leave for future work.

\subsection{\rva{\nbcr{4}{Applicability} of \X{} and its Defenses\\\nbcr{4}{to} Complex {PiM} Architectures}}
\rva{Our observations apply to \nbcr{4}{other} PiM architectures as long as they satisfy the key properties that enable the attack: 1) allowing co-locating data from different applications in a DRAM bank, and 2) enabling subsequent accesses from different applications targeting the same DRAM bank. 
Complex \gls{pnm} architectures (e.g., systems like Tesseract~\cite{Tesseract,he2025papi,gu2025pim} that employ sophisticated prefetching and network functionality) can degrade the attack throughput by introducing additional noise.}

\rva{The \nbcr{4}{performance overhead} of \nbcr{4}{an \X{}} defense heavily depends on how the defense mechanism interacts with the underlying system. The defense mechanisms we discuss in \secref{sec:mitigations} are not specifically tailored for \nbcr{4}{a given} underlying system (e.g., constant-time policy). The new defense mechanism that we introduce in \secref{sec:selective} enables the constant-time policy only for the bank under contention, \nbcr{4}{reducing the performance overhead}, but does not provide strong security guarantees.}

\subsection{\rvd{Restricting Access to PIM Operations}}\labeld{1}
\rvd{A key contribution of our study is to show that unrestricted access to PiM operations enables potentially harmful high-throughput side and covert channel attack vectors. None of the current PiM proposals \nbcr{4}{we are aware of} discuss restricting access to PiM operations to ensure security.
We hope and believe that our study will guide system designers in balancing the performance, flexibility, and security \nbcr{4}{trade-offs} while implementing \nbcr{4}{effective} access control \nbcr{4}{mechanisms to PiM}.}

\subsection{\rvd{\nbcr{4}{Applicability to Future DRAM Devices}}}

\rvd{The row buffer timing channel would persist in \nbcr{4}{future DRAM-based} systems, enabling IMPACT. \nbcr{4}{With newer generations, we observe that DRAM modules consist of increasingly more banks, which increases \X{}'s covert channel throughput as the attack would benefit from more parallelism. Another change is the introduction of RowHammer mitigations, PRAC~\cite{jedec2024jesd795c,canpolat2025chronus,canpolat2024understanding} and RFM~\cite{jedec2020jesd795}, in recent DDR specifications that block DRAM accesses when necessary to prevent RowHammer bitflips. These mitigations} incur at least 350ns (at most 1400ns) latency per preventive action. This is significantly higher than the row buffer conflict latency and can be filtered out by the receiver. 
}

\rvd{We believe that increasing DRAM’s complexity will increase attack vectors as manufacturers try to combat the challenges of technology scaling. For example, recent preprints show that \nbcr{4}{PRAC and RFM} introduce timing channels that can be exploited for side-channel~\cite{bostanci2025understanding} and covert-channel~\cite{bostanci2025understanding,taneja2025roguerfm} attacks.}

\section{Related Work}
\label{sec:relatedwork}
To our knowledge, this is the first work that evaluates \zb{timing attack vulnerabilities in PiM systems}. 
\sloppy
\subsection{{Security Implications of PiM Systems}} A set of prior works~\cite{pimsecchallengesGLSVLSI2020,ensan2021scare,Wang_2023} discuss specific security vulnerabilities in PiM systems. 
Arafin et al.~\cite{pimsecchallengesGLSVLSI2020} survey and theoretically analyze security vulnerabilities, focusing on five specific attacks. In this work, we quantitatively and qualitatively analyze and evaluate main memory-based covert and side channels in detail. Wang et al.~\cite{Wang_2023}  \nbcr{4}{extract} neural network architectural information from an RRAM accelerator using power side channels. Ensan et al.~\cite{ensan2021scare} \nbcr{4}{propose} an attack on RRAM-based PiM architectures using power/timing side channels. In contrast, we demonstrate how to establish high-throughput timing covert and side channels using PiM operations.

\subsection{Other Timing Attacks} 
\aedraft{We already discuss main memory-based timing attacks in~\secref{sec:mainmemorytiming}. In addition, prior works showcase multiple cache-based timing side and covert-channel attacks \nbcr{4}{(e.g., \cite{maurice2015c5, yan2016replayconfusion, kaur2023tppd, saileshwar2021streamline, gruss2016flush+, yarom2014flush+, xiong2020leaking,aciiccmez2007yet, xiong2021leaking})}. 
Among these works, Streamline \cite{saileshwar2021streamline} has the highest throughput as it avoids explicit cache flushes and utilizes a large transmission set to evict cache lines implicitly by thrashing the cache. \X{} achieves a leakage throughput within the same order of magnitude of Streamline and does \textit{not} require a transmission set larger than the cache size.
}

\noindent
\subsection{{Other Main Memory Timing Attack Defenses}}
We already extensively discuss defenses that fundamentally eliminate the timing channel (\secref{sec:mitigations}). % to PiM-based attacks. 
Another approach is to restrict the \nbcr{4}{high-resolution} timers' usage to prevent attackers from measuring memory requests' latencies, which is applied in some modern processors \atb{(e.g., Apple M1 cores~\cite{ravichandran2022pacman})}. However, these timers are used in many userspace applications (e.g., performance and power consumption monitoring, multithreading, \atb{and} synchronization and locking).
% , optimizations on protocols such as TCP. 
Restricting access to these timers can disable these applications or degrade their \atb{performance}.
\revct{DAGguise~\cite{deutsch2022dagguise}\revc{\label{resp:c1}1}\label{sec:dagguise} is a defense mechanism against memory timing side-channel attacks that obfuscates an application’s memory access pattern. 
DAGguise does not mitigate row buffer-based timing channels and requires a closed row policy.} 
Trusted execution environments (TEEs) (e.g., Intel SGX~\cite{anati2013innovative,mckeen2013innovative}) can isolate applications from the rest of the system. 
\nbcr{4}{Prior} works~\cite{moghimi2017cachezoom,gotzfried2017cache,schwarz2020malware,wang2017leaky,moghimi2019memjam,xu2015controlled,gyselinck2018off,van2017telling,lee2017inferring,evtyushkin2018branchscope} already \nbcr{4}{demonstrate} multiple side-channel attacks on TEEs leveraging main memory~\cite{wang2017leaky}, \nbcr{4}{cache hierarchy~\cite{moghimi2017cachezoom,gotzfried2017cache,schwarz2020malware,wang2017leaky,moghimi2019memjam}, page table entries~\cite{xu2015controlled,gyselinck2018off,van2017telling,wang2017leaky}, and branch predictors~\cite{lee2017inferring,evtyushkin2018branchscope}.}

\subsection{{Enhancing System Security using PiM}}

Several prior works enhance \nbcr{4}{system} security by leveraging PiM.
 \revet{SecNDP~\cite{secndpHPCA2022}\reve{1} proposes an encryption and verification mechanism for untrusted PiM devices.
 It only focuses on protecting confidentiality and does not consider or analyze potential timing channels.}
  Nelson et al.~\cite{nelson2022eliminating} propose offloading security-critical tasks to \nbcr{4}{\gls{pnm}} units to avoid storing sensitive information in caches and eliminate cache-based side-channel attacks. In this work, we demonstrate and evaluate the implications of PiM on main memory-based timing attacks.
InvisiMem~\cite{invisiMemISCA2017} and ObfusMem~\cite{awad2017obfusmem} propose leveraging \nb{3D-stacked memory} to design secure processors that provide oblivious RAM equivalent guarantees.

\section{Conclusion}

We introduce IMPACT, a set of high-throughput main memory timing attacks that leverage the direct memory access enabled by \nbcr{4}{Processing-in-Memory} (PiM) architectures. \nba{\X{} exploits the shared DRAM row buffer to 1) establish covert channels to communicate with other processes at high throughput and 2) leak security-critical information of victim processes based on their memory accesses. IMPACT covert channels leverage parallelism and eliminate cache bypassing steps, resulting in increased throughput compared to existing main memory-based covert channels. \X{} side-channel attack targets a genomics application and leaks the private information with a low error rate.} We evaluate and discuss four potential defense mechanisms against IMPACT and eventually conclude that mitigating \X{} incurs high performance overhead and more research is needed to find low-overhead solutions.

\section*{{Acknowledgments}}
{We thank the anonymous reviewers of USENIX Security'24, ISCA'24, MICRO'24, ASPLOS'25, and DSN'25 for their constructive feedback. 
We thank the SAFARI Research Group members for providing a stimulating intellectual environment. 
We acknowledge the generous gifts from our industrial partners, \nbcr{4}{including Google, Huawei, Intel, and Microsoft. This work, along with our broader work in Processing-in-Memory and memory systems, is supported in part by the Semiconductor Research Corporation (SRC), the ETH Future Computing Laboratory (EFCL), AI Chip Center for Emerging Smart Systems (ACCESS), sponsored by InnoHK funding, Hong Kong SAR, European Union’s Horizon programme for research and innovation [101047160 - BioPIM], a Google Security and Privacy Research Award, and the Microsoft Swiss Joint Research Center.}
}

%%%%%%% -- PAPER CONTENT ENDS -- %%%%%%%%
%\setstretch{1}
\balance
%%%%%%%%% -- BIB STYLE AND FILE -- %%%%%%%%
\bibliographystyle{IEEEtran}
\bibliography{refs}
%%%%%%%%%%%%%%%%%%%%%%%%%%%%%%%%%%%%

\clearpage
\appendix

\nobalance

\section{Artifact Appendix}

\subsection{Abstract}

This artifact provides the source code and scripts to reproduce the experiments and results presented in our DSN 2025 paper. It includes implementations of various covert channel attacks (PNM, DRAMA-clflush, DRAMA-evict, PUM, DMA) and a defense mechanism, along with scripts to automate the setup, execution, and result analysis.

\subsection{Artifact Checklist (Meta-information)}

{\small
\begin{itemize}
    \item {\bf Program:} C++ programs, Python scripts, shell scripts.
    \item {\bf Compilation:} GNU Make.
    \item {\bf Run-time environment: } Linux (tested on Ubuntu 20.04 and 22.04), Python 3.
    \item {\bf Execution: } Bash scripts, Python scripts.
    \item {\bf Metrics: } Timing results, performance metrics of attacks and defenses.
    \item {\bf Output: } Figures in PDF format and related data in plaintext files.
    \item {\bf How much disk space required (approximately)?: } 20GB.
    \item {\bf How much time is needed to prepare workflow (approximately)?: } $\sim1$ hour.
    \item {\bf How much time is needed to complete experiments (approximately)?:} 2-4  hours.
    \item {\bf Publicly available?: } Yes
    \item {\bf Archived (provide DOI)?: } Yes, DOI: 10.5281/zenodo.15116851
\end{itemize}
}

\subsection{Description}
\subsubsection{How to Access} The source code and scripts can be downloaded from Zenodo (\url{https://zenodo.org/records/15116851}) and GitHub (\url{https://github.com/CMU-SAFARI/IMPACT}).

\subsubsection{Hardware dependencies}
\begin{itemize}
\item x86-64 system.
\item At least 8GB of RAM (128GB recommended for parallel defense experiments).
\item $>64$ cores are suggested for parallel defense experiments.
\end{itemize}

\subsubsection{Software dependencies}
\begin{itemize}
    \item Linux Operating System (tested on Ubuntu 20.04 and 22.04).
    \item Bash.
    \item Python (version 3.12.9)
    \item GNU Make (version 4.3).
    \item wget, tar.
    \item GCC (11.4.0), g++ (11.4.0). 
    \item zlib1g-dev.
\end{itemize}

\subsubsection{Data sets}
The defense mechanism experiments may require trace files. The script includes commented-out lines to download example traces. You may need to download these or provide your own traces.

\subsection{Installation}

\subsubsection{Dependency Installation}
To install all dependencies, run the dependency installation script: \texttt{install\_dependencies.sh}.

\fancycommand{\$ cd IMPACT\\\$ sh install\_dependencies.sh\\\$ bash\\\$ conda activate impact}

\subsubsection{Conda Environment Setup}
If you would like to install Miniconda manually, follow the steps below.

\begin{enumerate}
    \item Install Miniconda from \url{https://docs.conda.io/en/latest/miniconda.html}.
    \item Run the Miniconda installer.
    \item Open a new terminal.
\end{enumerate}

\subsection{Experiment Workflow}
To run all experiments, run the main script: \texttt{artifact.sh}. This script 1) builds the attacks and the simulator and 2) runs all experiments required to reproduce the key results.

\fancycommand{\$ sh artifact.sh}

The \texttt{artifact.sh} script will execute the following steps.

\subsubsection{Building the Attacks and Simulator}
\begin{enumerate}
    \item The script will navigate to the \texttt{impact/covert\_channel\_attack/} directory.
    \item It will clean and build the attack code using \texttt{make clean} and \texttt{make all}.
    \item It will navigate to the \texttt{simulator/sniper} directory.
    \item It will clean and build the simulator using \texttt{make distclean} and \texttt{make -j\$(nproc)}.
\end{enumerate}

\subsubsection{Running Examples and Attacks}

\subsubsection{Basic Example Run}
The script includes a commented-out line to run a basic example using the Sniper simulator. Uncomment the line.

\subsubsection{Running Attacks for Figure 9}
The script executes the following attacks:
\begin{itemize}
    \item PNM: \texttt{python3 \$home\_dir/impact/covert\_channel\_attack/script/run\_pnm.py \$home\_dir 16 \$home\_dir/results\_attacks/pnm}
    \item DRAMA (clflush): \texttt{python3 \$home\_dir/impact/covert\_channel\_attack/script/run\_drama.py \$home\_dir 16 \$home\_dir/results\_attacks/drama}
    \item DRAMA (eviction sets): \texttt{python3 \$home\_dir/impact/covert\_channel\_attack/script/run\_drama\_ev.py \$home\_dir 16 \$home\_dir/results\_attacks/drama\_ev}
    \item PUM: \texttt{python3 \$home\_dir/impact/covert\_channel\_attack/script/run\_pum.py \$home\_dir 16 \$home\_dir/results\_attacks/pum}
    \item DMA: \texttt{python3 \$home\_dir/impact/covert\_channel\_attack/script/run\_dma.py \$home\_dir 16 \$home\_dir/results\_attacks/dma}
\end{itemize}

\subsubsection{Running Attacks for Figure 8 and 10 (PNM POC)}
\texttt{python3 \$home\_dir/impact/covert\_channel\_attack/script/run\_pnm\_poc.py \$home\_dir 16 \$home\_dir/results\_attacks/pnm\_poc}

\subsubsection{Running Attacks for Figure 8 and 10 (PUM POC)}
\texttt{python3 \$home\_dir/impact/covert\_channel\_attack/script/run\_pum\_poc.py \$home\_dir 16 \$home\_dir/results\_attacks/pum\_poc}

\subsubsection{Running Defense Mechanism Experiments for Figure 12}
\begin{itemize}
    \item Download traces (if needed).
    \item Create the jobfile: \texttt{python3 \$home\_dir/scripts/create\_jobfile\_impact.py \$home\_dir/simulator/sniper \$home\_dir/scripts/defenses.jobfile}
    \item Execute the jobfile: \texttt{sh \$home\_dir/scripts/defenses.jobfile}
\end{itemize}

\subsection{Plotting the Results}

Once the experiments are completed, the reader can plot the figures using the plotting scripts we provide.

To reproduce the figures, 1) install plotting dependencies:

\fancycommand{\$ pip install numpy matplotlib pandas}

2) run the plotting script: \texttt{plot\_figures.sh}

\fancycommand{\$ sh plot\_figures.sh}

This script will create Figures 8, 9, 10, and 12 under the \texttt{figures/} directory.

\subsection{Evaluation \& Expected Results}

\subsubsection{Post-Execution Steps}
Review Figures 8, 9, 10, and 12 in the \texttt{figures/} directories.

Note that the attack codes are compiled during the setup, and the results presented in Figures 8, 9, and 10 might vary slightly based on your system configurations (e.g., compiler version). 

\subsubsection{Troubleshooting}
Refer to the troubleshooting section in the README.md for potential issues and solutions.

\subsection{Methodology}

Submission, reviewing and badging methodology:

\begin{itemize}
    \item \url{https://www.acm.org/publications/policies/artifact-review-and-badging-current}
    \item \url{http://cTuning.org/ae/submission-20201122.html}
    \item \url{http://cTuning.org/ae/reviewing-20201122.html}
\end{itemize}

\end{document}